\documentclass[prx,onecolumn,floatfix,longbibliography,notitlepage, nofootinbib]{revtex4-1}

\usepackage[letterpaper,top=2cm,bottom=2cm,left=3cm,right=3cm,marginparwidth=1.75cm]{geometry}

\usepackage{tcolorbox}
\tcbuselibrary{breakable}
\usepackage{graphicx, color, graphpap}
\usepackage{subcaption}
\usepackage{enumitem}
\usepackage{amssymb}
\usepackage{amsthm}
\usepackage{dsfont}
\usepackage{mathtools}
\usepackage{multirow}
\usepackage[colorlinks=true,citecolor=magenta,linkcolor=blue]{hyperref}
\usepackage[T1]{fontenc}
\usepackage{bbm}
\usepackage{thmtools,thm-restate}
\usepackage{verbatim}
\usepackage{mathtools}
\usepackage{titlesec}
\usepackage{amsmath}
\usepackage[normalem]{ulem}
\usepackage{circuitikz}
\usepackage{braket}
\usepackage{ragged2e}

\usepackage{listings}
\usepackage{csquotes}

\newcommand{\dif}{\mathrm{d}} 

\long\def\ca#1\cb{} 

\begin{document}

\title{Thermodynamic Computing System for AI Applications}
\author{Denis Melanson, Mohammad Abu Khater, Maxwell~Aifer, Kaelan~Donatella, Max~Hunter~Gordon, Thomas~Ahle, Gavin Crooks, Antonio J. Martinez, Faris Sbahi, Patrick J. Coles}

\affiliation{Normal Computing Corporation, New York, New York, USA}

\begin{abstract}
Recent breakthroughs in artificial intelligence (AI) algorithms have highlighted the need for novel computing hardware in order to truly unlock the potential for AI. Physics-based hardware, such as thermodynamic computing, has the potential to provide a fast, low-power means to accelerate AI primitives, especially generative AI and probabilistic AI. In this work, we present the first continuous-variable thermodynamic computer, which we call the stochastic processing unit (SPU). Our SPU is composed of RLC circuits, as unit cells, on a printed circuit board, with 8 unit cells that are all-to-all coupled via switched capacitances. It can be used for either sampling or linear algebra primitives, and we demonstrate Gaussian sampling and matrix inversion on our hardware. The latter represents the first thermodynamic linear algebra experiment. We also illustrate the applicability of the SPU to uncertainty quantification for neural network classification. We envision that this hardware, when scaled up in size, will have significant impact on accelerating various probabilistic AI applications.
\end{abstract}

\maketitle

\section{Introduction}

The AI revolution has highlighted the shortcomings of today's computing hardware. AI leaders have argued~\cite{hooker2021hardware} that machine learning is currently stuck in a local optimum, and if the field could move away from current digital hardware, then a global optimum could be reached. Probabilistic AI, in particular, which is a branch of AI dealing with Bayesian inference, uncertainty quantification, and sampling tasks and has led to recent breakthroughs in generative AI like diffusion models, has been noted to struggle with computational difficulty on current digital hardware~\cite{izmailov2021bayesian,LeCun2023MITphysics}. This has spawned calls for mortal computation~\cite{HintonNeurIPS2022}, where software and hardware are inseparable and the hardware is variable and stochastic.

Analog computing offers an appealing alternative to today's digital computers, both in terms of energy efficiency, and also potentially in terms of processing speed, if one can match the physics of the analog hardware to the mathematics of the AI algorithms. Optimization has been a focus of several analog, physics-based computing demonstrations~\cite{mohseni2022ising,aadit2022massively,mourgias2023analog,inagaki2016coherent,moy20221,chou2019analog,wang2019oim,theilman2023stochastic}. We argue that a more natural match between the physics of the analog hardware and the mathematical application may come from considering probabilistic AI. For this application, the mathematics happens to match that of thermodynamics~\cite{coles2023thermodynamic}, which is the branch of classical physics that involves stochastic dynamics.

The relevance of thermodynamics to solving mathematical problems has recently spawned the field of thermodynamic computing~\cite{conte2019thermodynamic}, leading to several hardware proposals~\cite{coles2023thermodynamic,aifer2023thermodynamic,duffield2023thermodynamic,hylton2020thermodynamic,ganesh2017thermodynamic,lipka2023thermodynamic,Camsari_2019,chowdhury2023full,misra2023probabilistic,liu2022bayesian,mansinghka2009natively} including some with application to accelerating probabilistic AI, such as discrete-variable hardware based on probabilistic bits~\cite{Camsari_2019,chowdhury2023full,misra2023probabilistic,liu2022bayesian,mansinghka2009natively} and continuous-variable hardware based on analog circuits~\cite{coles2023thermodynamic,aifer2023thermodynamic,duffield2023thermodynamic}. Thermodynamic computers may have inherent robustness to noise since noise is actually a desirable feature of the hardware~\cite{coles2023thermodynamic}.

Continuous-variable (CV) thermodynamic computers are especially well suited for probabilistic AI, since Bayesian inference typically involves continuous probability distributions, such as Gaussians or the exponential family~\cite{murphy2022probabilistic}. Moreover, generative AI methods such as diffusion models involve continuous-variable Brownian motion, which is once again natural to simulate on CV thermodynamic computers. CV thermodynamic computers also have application to the central mathematical primitives in AI, namely, linear algebra~\cite{aifer2023thermodynamic,duffield2023thermodynamic}, such as solving linear systems, inverting matrices, and computing matrix determinants and matrix exponentials. Hence, it is natural to consider the CV version of thermodynamic computers for AI applications.

In this work, we present the first CV thermodynamic computer. Our device is fabricated as a printed circuit board, with 8 fully-connected unit cells. We experimentally demonstrate that this computer can be used either as a sampling device, to sample from user-specified Gaussian distributions, or as a linear algebra device, to invert user-specified matrices. The latter represents the first experimental implementation of a thermodynamic linear algebra experiment. We also perform demonstrations on our hardware of important primitives in Probabilistic AI~\cite{murphy2022probabilistic}, including Gaussian Process Regression (GPR)~\cite{williams1995gaussian} and Spectral Normalized Neural Gaussian Processes (SNGP)~\cite{liu2020simple} for uncertainty quantification in neural networks. 

Our proof-of-principle demonstration highlights a future where thermodynamic advantage may be a reality, i.e., where thermodynamic computers outperform digital ones in either speed or energy efficiency. Indeed, asymptotic speedups have been been theoretically predicted~\cite{aifer2023thermodynamic,duffield2023thermodynamic}, implying that there exists a threshold scale (or problem size) beyond which thermodynamic computers will outperform competition. Numerical simulations furthermore suggest that this threshold scale is practically achievable~\cite{aifer2023thermodynamic}, and we provide further evidence to this claim herein. Finally, we note that, while one path to thermodynamic advantage is to directly scale up the hardware design presented here (for Gaussian sampling and linear algebra), small perturbations to our hardware design could unlock other applications, such as non-Gaussian sampling or generative diffusion models~\cite{coles2023thermodynamic}, with additional opportunities for thermodynamic advantage.

\section{Results}

\subsection{The Computing System}\label{sec3}

\subsubsection{Fundamentals of thermodynamic computing}

We begin by reviewing thermodynamic computing, which is a relatively young field, and hence the theoretical concepts are still being worked out. We focus here on the paradigm of CV thermodynamic computing from Refs.~\cite{coles2023thermodynamic,aifer2023thermodynamic,duffield2023thermodynamic}, which is based on the stochastic dynamics of a physical system acted on by a combination of conservative, dissipative, and fluctuating forces. These dynamics can be modeled by the underdamped Langevin (UDL) stochastic differential equations (SDEs)
\begin{align}
\label{eq:UDL-x}
    \dif x &= M^{-1} p \, \dif t\\
\label{eq:UDL-p}
\dif p &= -\nabla U - \gamma M^{-1} p \, \dif t + \mathcal{N}[0, 2 \gamma \beta^{-1}\mathbb{I} \, \dif t],
\end{align}
where $\gamma$ and $\beta$ are positive real constants and $M$ may be either a positive real scalar or a positive definite matrix. The vectors $x$ and $p$ represent (respectively) generalized coordinates describing the system's state and their canonically conjugate momenta, and the function $U(x)$ is the potential energy, which is responsible for conservative forces. In our work, the quantities $x$ and $p$ will be mapped to (respectively) the currents and voltages in a circuit. Also, the noise term here is assumed to have a covariance matrix proportional to identity (i.e., the noise is uncorrelated), but in general we may also replace the identity matrix by another positive definite matrix to model a system coupled to a correlated noise source. Note that Eqs. \eqref{eq:UDL-x} and \eqref{eq:UDL-p} can be combined to yield $\dif x = -\gamma^{-1} \nabla U - \gamma^{-1} \dif p + \mathcal{N}[0,2\gamma^{-1}\beta^{-1}\mathbb{I}\, \dif t]$; in the overdamped limit (where $M$ is small and $\gamma$ is large) the second term on the right can be dropped, leading to the overdamped Langevin (ODL) equation
\begin{equation}
    \label{eq:odl}
    \dif x = -\gamma^{-1}\nabla U + \mathcal{N}[0, 2\gamma^{-1}\beta^{-1} \mathbb{I} \, \dif t].
\end{equation}

Langevin equations (overdamped or underdamped) are examples of SDEs, which describe a system on the level of a single trajectory. In general, the SDE that governs the stochastic trajectories of a system can be recast as a partial differential equation (PDE) that governs the time-evolution of the probability density function of the system's state. The PDEs corresponding to the overdamped and underdamped Langevin equations are called Fokker-Planck equations, and the stationary distribution for $x$ (which is the same for the ODL and UDL equations) is called the Gibbs distribution
\begin{equation}
    f(x) = \frac{1}{Z}e^{-\beta U(x)},
\end{equation}
where the partition function $Z$ is determined by normalization~\cite{chen2014stochastic}. Clearly when $U(x)$ is a quadratic function $U(x) = x^T A x + b^T x$, the Gibbs distribution is Gaussian; specifically, it is the distribution
\begin{equation}
    x\sim \mathcal{N}[A^{-1} b, \beta^{-1}A^{-1}],
\end{equation}
which suggests thermodynamic algorithms for solving linear systems of equations and inverting matrices \cite{aifer2023thermodynamic}. This also allows for drawing samples from a Gaussian distribution by preparing the appropriate Gibbs state. However, these methods rely on the properties of the stationary distribution of a system, meaning the system must be allowed to come to equilibrium before useful samples can be read from the device. This time period is called the equilibration (or burn-in) time. Also, if two samples are drawn of the system's state with a very small interval of time in between, the values will be very similar, which is due to the time-correlation of the system. If uncorrelated samples are desired (which would be the case for i.i.d. Gaussian sampling tasks, for example), the timescale one must wait is approximately set by the \emph{correlation time} of the system. The equilibration time and correlation time place important constraints on the performance of thermodynamic algorithms.

\subsubsection{The Stochastic Processing Unit}

\begin{figure}[t]%
\centering
\includegraphics[width=0.83\textwidth]{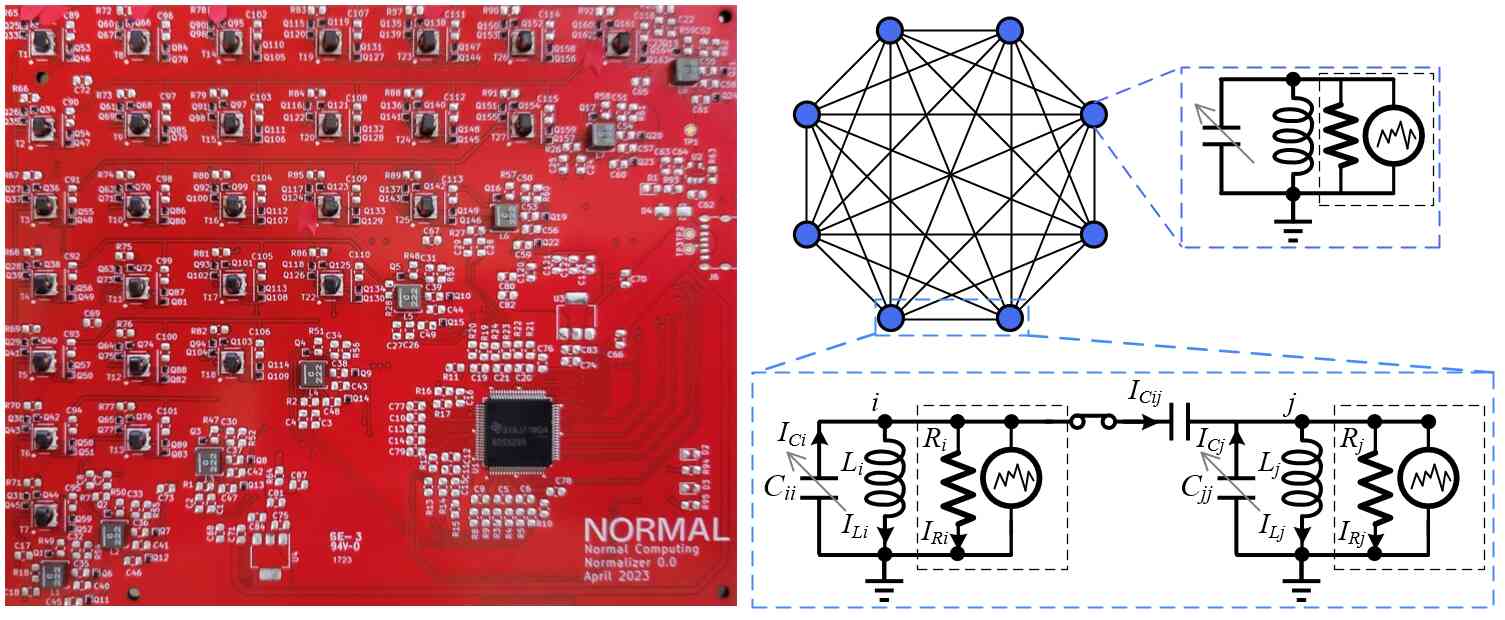}
\caption{\justifying \textbf{The Stochastic Processing Unit (SPU).} (Left panel) The Printed Circuit Board for our 8-cell SPU. (Right panel) Illustration of eight unit cells that are all-to-all coupled to each other, as in our SPU. Each cell contains an LC resonator and a Gaussian current noise source, as shown in the circuit diagram on the top right. The circuit diagram on the bottom depicts two capacitively coupled unit cells.}\label{fig:PCB_AllAll}
\end{figure}

We now introduce our stochastic processing unit (SPU), which is depicted in the left panel of Fig.~\ref{fig:PCB_AllAll}. The SPU is constructed on a Printed Circuit Board (PCB). From the lower left corner to the upper right corner, one can see the line of components corresponding to 8 unit cells (LC circuits), while the components arranged in the triangle on the upper left correspond to the controllable couplings that couple the unit cells. We remark that we constructed three nominally identical copies of our SPU circuit, to test the scientific reproducibility of our experimental results.

The SPU can be mathematically modeled as a set of capacitively-coupled ideal LC circuits with noisy current driving. The diagram for this model is shown in the right panel of Fig.~\ref{fig:PCB_AllAll}. Doing a simple circuit analysis reveals that the equations of motion for this circuit are
\begin{align}\label{eq:two_coupled_cells}
    \dif I &= \mathbf{L}^{-1} V \dif t \\
    \label{eq:langevin-V}
    \dif V &= -\mathbf{C}^{-1} \mathbf{R}^{-1} V \dif t - \mathbf{C}^{-1} I \dif t + \sqrt{2\kappa_0}\mathbf{C}^{-1} \mathcal{N}[0,\mathbb{I}\,\dif t],
\end{align}
where $I = \left(I_{L1},\dots I_{Ld}\right)^\mathrm{T}$ is the vector of inductor currents and $V = \left
(V_{C1}, \dots V_{Cd}\right)^\mathrm{T}$ is the vector of capacitor voltages. In the above, $\bold{C}$ is the Maxwell capacitance matrix, whose diagonal elements are $\bold{C}_{ii} = C_{ii} + \sum_{j=1}^dC_{ij}$, and whose off-diagonal elements are $\bold{C}_{ij} = -C_{ij}$. The values of resistors and inductors in each cell are represented by the matrices $\bold{R} = R \mathbb{I}$ and $\bold{L} = L \mathbb{I}$ respectively. Finally, $ \mathcal{N}[0,\mathbb{I} \,\dif t]$ represents a mean-zero normally distributed random displacement with covariance matrix $\mathbb{I}\, \dif t$ and $\kappa_0$ is the power spectral density of the current noise source. If the magnitude of the noisy driving current is larger than the intrinsic noise in the system, then $\kappa_0$ can be thought of as an effective temperature control for the thermodynamic computation.

Equations \eqref{eq:two_coupled_cells} and \eqref{eq:langevin-V} can be mapped to the Langevin equations \eqref{eq:UDL-x} and \eqref{eq:UDL-p} by making a change of coordinates. Specifically, we introduce the magnetic flux vector $\Phi$ and the Maxwell charge vector $\mathcal{Q}$, defined as
\begin{equation}
    \Phi = L I,  \: \: \: \: \: \: \: \mathcal{Q} = \bold{C} V.
\end{equation}
As  shown in the Supplemental Information, $\Phi$ and $\mathcal{Q}$ are canonically conjugate coordinates, with $\Phi$ playing the role of position and $\mathcal{Q}$ playing the role of momentum. We also introduce an effective inverse temperature parameter $\beta = \gamma \kappa_0^{-1}$. In terms of these variables, Eqs. \eqref{eq:two_coupled_cells} and \eqref{eq:langevin-V} become
\begin{align}
\label{eq:langevin-phi}
    \dif \Phi &= \bold{C}^{-1}\mathcal{Q} \, \dif t \\
    \label{eq:langevin-Q}
    \dif \mathcal{Q} &= -  L^{-1}\Phi\, \dif t - R^{-1} \bold{C}^{-1}\mathcal{Q}\, \dif t + \mathcal{N}[0,2R^{-1}\beta^{-1}\mathbb{I}\,\dif t].
\end{align}
It is clear that Eqs.~\eqref{eq:langevin-phi} and \eqref{eq:langevin-Q} are equivalent to \eqref{eq:UDL-x} and \eqref{eq:UDL-p} when we make the identifications $x = \Phi$, $p = \mathcal{Q}$, $M = \bold{C}$, $\gamma = R^{-1}$, and $U(x) = U\left(\Phi\right) = \frac{1}{2}\Phi^T \bold{L}^{-1} \Phi$. In these coordinates the Hamiltonian, without noise or dissipation, of the system is expressed as
\begin{align}
    \mathcal{H} \left(\Phi, \mathcal{Q}\right) &=   \frac{1}{2}\Phi^T \mathbf{L}^{-1} \Phi + \frac{1}{2}\mathcal{Q}^T\mathbf{C}^{-1}\mathcal{Q},
\end{align}
and consequently the stationary distribution of Eqs.~\eqref{eq:langevin-phi} and \eqref{eq:langevin-Q} is the Gibbs distribution given by
\begin{equation}
    \label{eq:gibbs-Q-Phi}
    \Phi \sim \mathcal{N}[0,\beta^{-1}\bold{L}], \: \: \: \: \: \: \mathcal{Q} \sim \mathcal{N}[0,\beta^{-1}\bold{C}],
\end{equation}
where $\Phi$ and $\mathcal{Q}$ are independent of each other.

\begin{figure}[t]
    \centering
    \includegraphics[width=0.55\columnwidth]{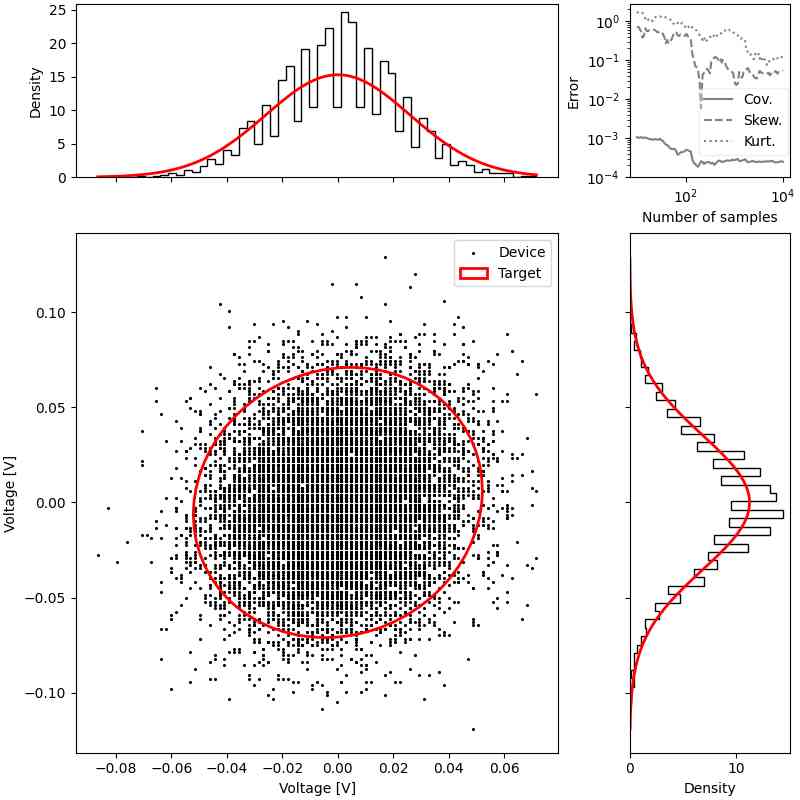}
    \caption{\justifying \textbf{Voltage samples from two coupled unit cells of the SPU.} Top Left: Histogram of the marginal of cell $i$. Top Right: Absolute error between the target covariance matrix and the device covariance matrix, similarly for the skewness and kurtosis, all calculated using the Frobenius norm. Bottom Left: Scatter plot of voltage samples from both cells. Bottom Right: Histogram of the marginal of cell $j$. For the marginal plots, the theoretical target marginal is overlaid as a solid red curve. Similarly, for the two-dimensional plot, the theoretical curve corresponding to two standard deviations from the mean is overlaid as a solid red curve.}
    \label{fig:two_dim_sampling}
\end{figure}

This equilibrium distribution is reached after a sufficient amount of time has elapsed, called the equilibration time. The equilibration time is closely related to the correlation time $\tau_\text{corr}$, which is the timescale over which the time-correlation function of the system decays. In fact, equilibration can be interpreted as the decorrelation of the system from its initial state, so the two timescales are essentially the same. The correlation function decays exponentially in time with a time constant  of approximately
\begin{equation}
\tau_\text{corr} \approx R c_\text{max},
\end{equation}
where $c_\text{max}$ is the largest eigenvalue of $\bold{C}$ (see e.g. \cite{aifer2023thermodynamic}). There are some minor corrections involving the other circuit parameters but these have relatively little effect. The amount of time one must wait in between samples is determined by the degree to which correlation must be suppressed. In order for the correlation function to decay to less than one percent of its original magnitude, we may wait for an interval of at least $5 \tau_\text{corr}$, for example.

If one periodically measures the voltages, $V$, across the capacitors after the device has reached equilibrium, one finds that the voltage samples will have a covariance matrix of
\begin{align}
    \Sigma_V = R \kappa_0 \mathbf{C}^{-1}.
\end{align}
Figure~\ref{fig:two_dim_sampling} is a visualization of such an experiment, where the measurements of the voltages of two coupled cells are taken at 12 MHz. 
However, the sampling rate that one chooses for these measurements is an important consideration. Sampling too fast will decrease the fidelity of the samples to the distribution due to the errors from their time-correlation.

\subsubsection{Adjustable hardware parameters}

The particular implementation of a CV thermodynamic computer reported in this work is composed of 8 of the unit cells previously discussed with all-to-all connectivity. To increase the domain of addressable 8-dimensional Gaussian distributions, we designed each cell to have a switchable capacitor bank instead of a single capacitor. These capacitor banks have four distinct values of effective capacitance and as such give us 4 possible variances (diagonal elements of the covariance matrix), labeled as 0, 1, 2, 3. The capacitive couplings responsible for the covariances between voltages are also switchable and bipolar. Each of the 28 coupling elements is made of a single capacitor and a transformer. The transformer is present to make each coupling bipolar (i.e., having either positive or negative sign). It has two outputs, one that keeps the polarity of the signal and one that inverts it, each having a winding ratio of 1:1. The coupling can also be switched off, this make for three possible values of the effective coupling, labeled as -1, 0, and +1.

\begin{figure}[t]
    \centering
    \includegraphics[width=0.45\textwidth]{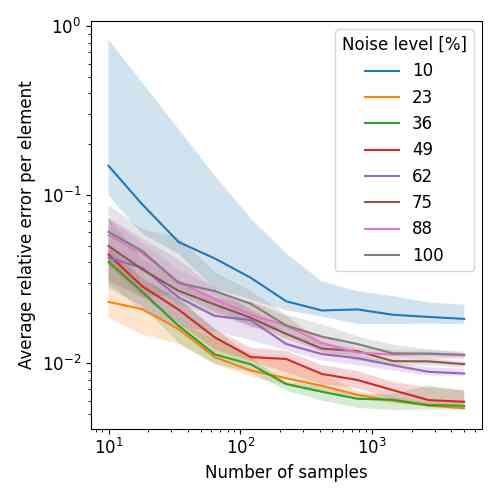}
    \caption{\justifying \textbf{Effect of noise level and number of samples on sample quality.} The noise level represents the amplitude of the stochastic noise injected into the unit cells. The $y$-axis plots the error on the covariance matrix, namely the average relative Frobenius error per matrix element. Samples are taken from the SPU with all positive couplings turned on and with the unit cell capacitances in configuration 3.}
    \label{fig:noise_level}
\end{figure}

In addition to the set of possible capacitances, two other key parameters can be varied: the noise level and the sampling rate. The noise level, $\kappa_0$, refers to the amplitude of the stochastic noise that is injected into each unit cell. In the SPU, the noisy current source is implemented using pseudo-random Gaussian currents from a Field-Programmable Gate Array (FPGA, see Methods). The amplitude of the noise current can be controlled and it can be viewed as an effective temperature, namely the temperature of a thermal resistor in each unit cell. Figure~\ref{fig:noise_level} shows the effect of noise level on the performance. One can see that intermediate noise levels appear to perform well, while noise levels that are too high or too low can lead to worse performance. This behavior could be due to the non-ideal behavior of the pseudo-random noise source at low or high amplitudes.

\begin{figure}[t]
    \centering
    \includegraphics[width=0.65\textwidth]{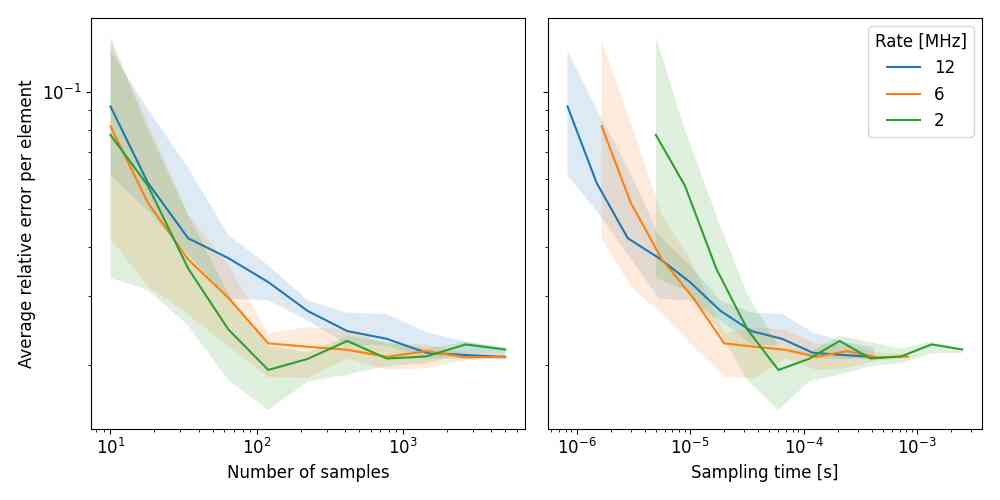}
    \caption{\justifying \textbf{Effect of sampling rate and number of samples on sample quality.} The $y$-axis plots the error on the covariance matrix, namely the average relative Frobenius error per matrix element. The left panel varies the number of samples, while the right panel varies the sampling time (i.e., the total length of time over which one draws samples). Samples are taken from the SPU with all positive couplings turned on and with the unit cell capacitances in configuration 3.}
    \label{fig:sampling_rate}
\end{figure}

The sampling rate refers to the frequency at which the hardware gathers the voltage samples from the unit cells, i.e., measures the state variable. In this implementation, we kept the base sampling frequency of the hardware fixed at 12 MHz and did down-sampling digitally in post-processing to investigate the effect of this parameter. Figure~\ref{fig:sampling_rate} shows the effect of sampling rate on the performance, for a range of sampling rates from 2 MHz to 12 MHz. One can see that the highest rate, 12 MHz, performs slightly better than slower rates if one's goal is to minimize overall sampling time. On the other hand, lower sampling rates perform better if one is concerned with the error for a certain number of samples.

\subsection{Gaussian sampling}

Let us describe how to perform Gaussian sampling with our thermodynamic computer. Consider a zero-mean multivariate Gaussian distribution (since we can always translate the samples by a constant vector to generate a non-zero mean):
\begin{align}\label{eq:multi_gaussian}
    \mathcal{N}(\vec{x}|\mathbf{\Sigma}) = \frac{1}{\sqrt{(2\pi)^N |\mathbf{\Sigma}|}}\exp\left(-\frac{1}{2}\vec{x}^T\mathbf{\Sigma}^{-1}\vec{x}\right),
\end{align}
where $\Sigma$ is the covariance matrix. Here we consider the case where the user provides the precision matrix $\mathbf{P}= \mathbf{\Sigma}^{-1}$ associated with the desired Gaussian distribution (See Supplemental Information for the alternative case where the user provides the covariance matrix~$\mathbf{\Sigma}$.)

The Hamiltonian for the coupled oscillator system (see Supplemental Information for details) is given by:
\begin{align}\label{eq:ham_hmc_iv}
    \mathcal{H} \left(\vec{I}, \vec{V}\right) &= \frac{1}{2}\vec{V}^T\mathbf{C}\vec{V} + \frac{1}{2}\vec{I}^T \mathbf{L} \vec{I},
\end{align}
where $\vec{I}$ is the vector of currents through the inductors in each unit cell, $\vec{V}$ is the vector of voltages across the capacitors in each unit cell, $\mathbf{C}$ is the Maxwell capacitance matrix and $\mathbf{L}$ is the inductance matrix, respectively given by
\begin{align}
    \mathbf{C}_{kl} = 
    \begin{cases}
        \sum_j C_{kj} & \text{if } k=l \\
        - C_{kl} &  \text{if } k\neq l \\
    \end{cases},\quad\text{and}\quad
    \mathbf{L}_{kl} = 
    \begin{cases}
        L_{k} & \text{if } k=l \\
        0 &  \text{if } k\neq l \\
    \end{cases}.
\end{align}
Here, $C_{kk}$ and $L_{k}$ are the capacitance and inductance, respectively, of the $k$-th unit cell, and $C_{kj}$ for $j\neq k$ is the capacitance that couples the $k$-th and $j$-th unit cells.

At thermal equilibrium, the dynamical variables are distributed according to a Boltzmann distribution, proportional to $\exp (-\mathcal{H} / kT)$, and hence $\vec{V}$ is normally distributed according to:
\begin{equation}
\label{eqn:VoltageDistribution}
\vec{V} \sim \mathcal{N}[\vec{0},kT\mathbf{C}^{-1}]
\end{equation}
Thus, if the user specifies the precision matrix $\mathbf{P}$, then we can obtain the correct distribution for $\vec{V}$ by choosing the Maxwell capacitance matrix to be: \begin{equation}
\label{eqn:CkTP}
\mathbf{C} = kT \hspace{1pt}\mathbf{P} 
\end{equation}
Hence, this describes how we can map the user-specified matrix to the matrix of electrical component values, to obtain the desired distribution.

Figure~\ref{fig:two_dim_sampling} shows an example where we sample from an 8-dimensional Gaussian distribution on our SPU, for a particular user-specified precision matrix. One can see that the error on the covariance goes down fairly mononotically with the number of samples, while errors on the skewness and kurtosis shows some initial non-monotonic behaviour followed by a steady decline with increasing samples. Each of the 8 one-dimensional marginal distributions agree well with the theoretical distributions, suggesting that the SPU is sampling the correct distribution.

\begin{figure}[t]
    \centering
    \includegraphics[width=0.6\textwidth]{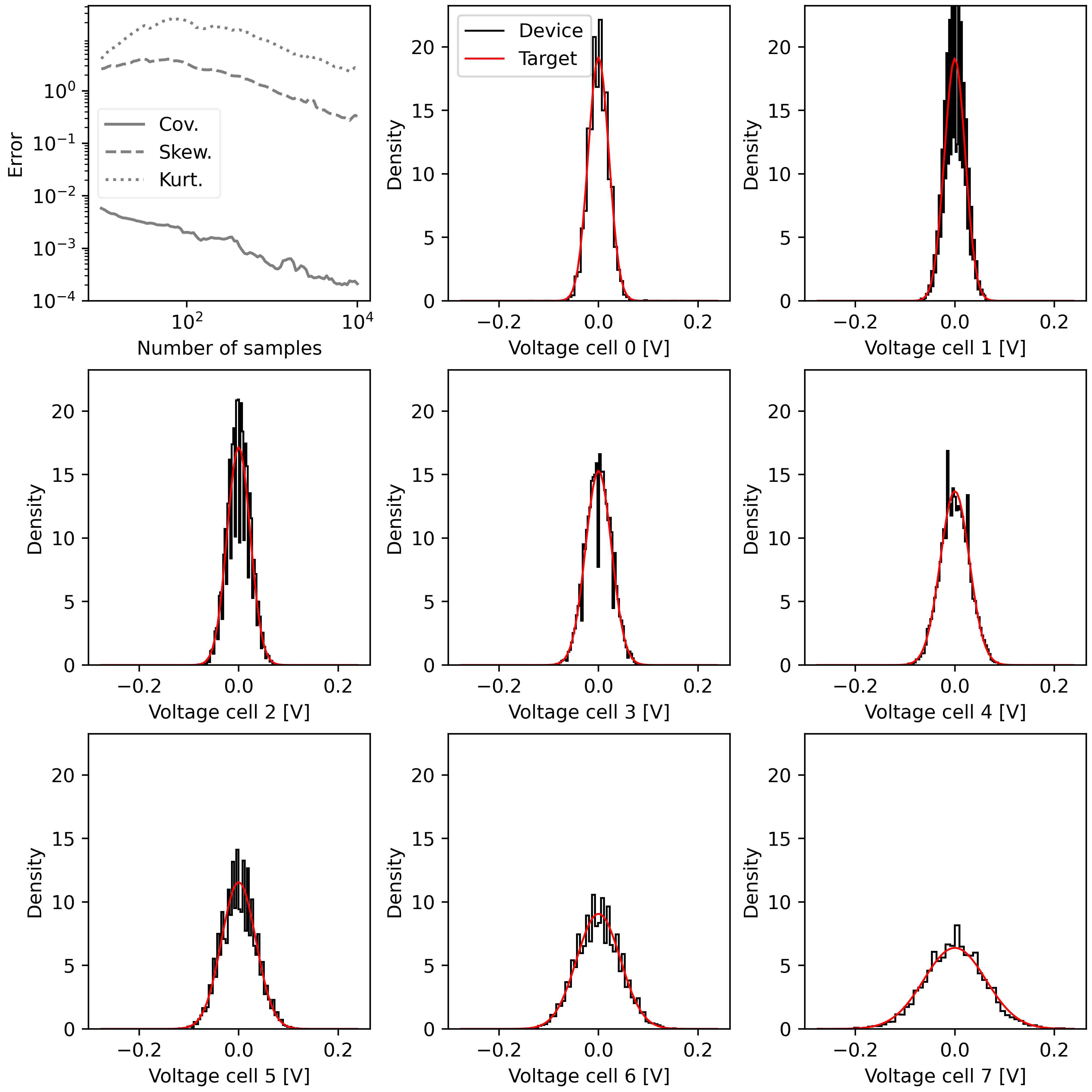}
    \caption{\justifying \textbf{Sampling from an 8-dimensional Gaussian distribution on the SPU.} The top left plot shows three error metrics, the error on the covariance, skewness, and kurtosis versus the number of samples. The error is the Frobenius norm of the difference between the target and device values. The other plots show the probability density versus voltage for each of the 8 one-dimensional marginal distributions. The experimentally determined histograms (in black) are overlaid with the target marginal distributions (in red). Samples are taken from the SPU with all couplings turned off and the unit cell capacitances are in configuration 3.}
    \label{fig:all_marginals}
\end{figure}

\subsection{Matrix inversion}

\begin{figure}[t]%
\centering
\includegraphics[width=0.55\textwidth]{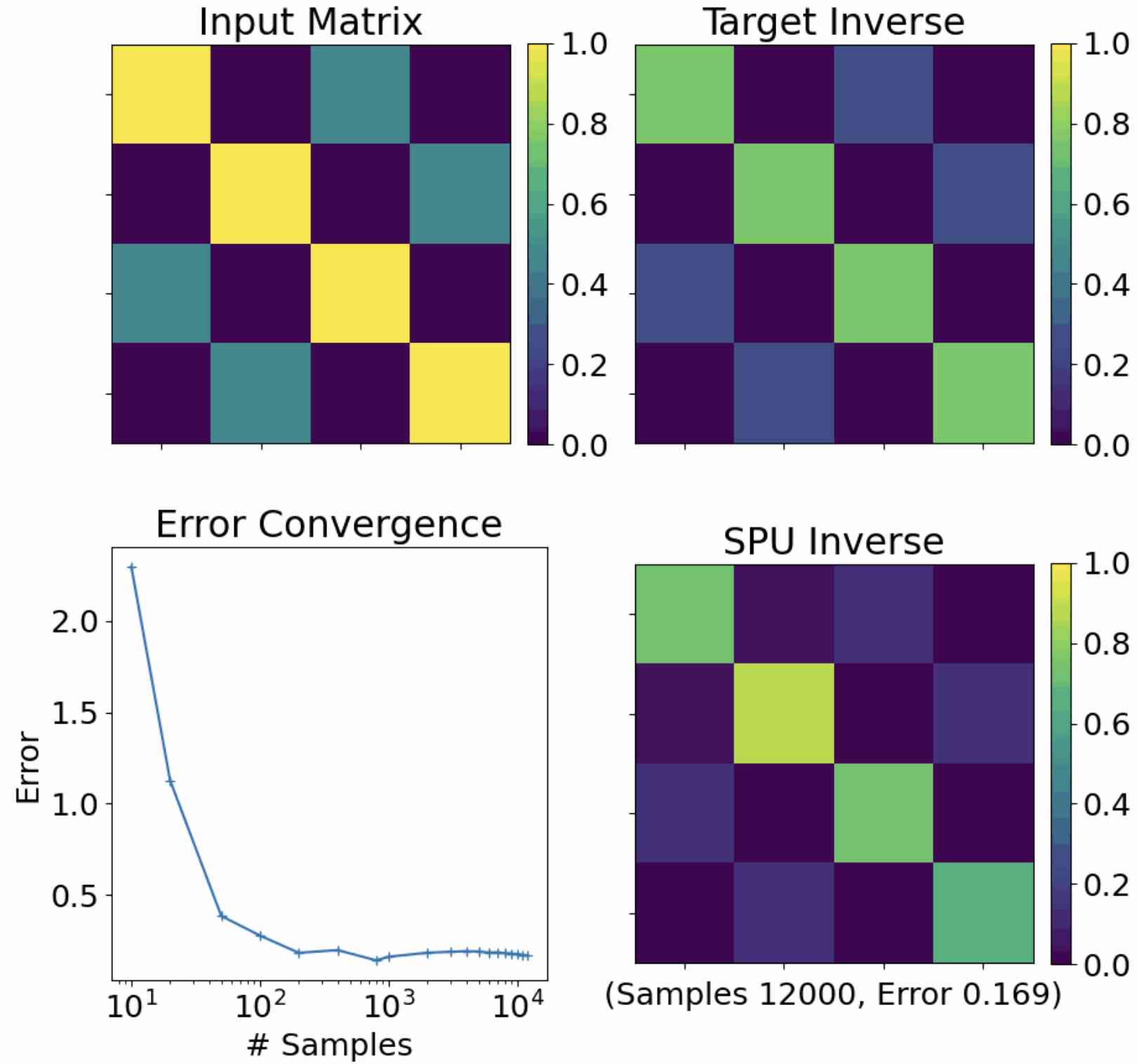}
\caption{\justifying \textbf{Thermodynamic inversion of a 4x4 matrix.} This experiment involved using only a subset of the unit cells, namely four of them, on the SPU. The input matrix $A$ and its true inverse $A^{-1}$ are shown, respectively, on the top left and top right. The relative Frobenius error versus the number of samples is plotted in the bottom left. The bottom right shows the experimentally determined inverse after gathering 12000 samples from the SPU. }\label{fig:Inversion4x4}
\end{figure}

The second primitive we will consider is matrix inversion, which was discussed in the context of thermodynamic computing in Ref.~\cite{aifer2023thermodynamic}. Following that reference, we envision the user encoding their matrix $\mathbf{A}$ in the precision matrix $\mathbf{P}$ of the associated Gaussian distribution that will be sampled. Hence from Eq.~\eqref{eqn:CkTP}, the Maxwell capacitance matrix of the hardware is given by $\mathbf{C} = kT\hspace{1pt} \mathbf{A}$. Choosing this Maxwell capacitance matrix, we find from Eq.~\eqref{eqn:VoltageDistribution} that at thermal equilibrium, the voltage vector is distributed according to 
\begin{equation}
\vec{V} \sim \mathcal{N}[\vec{0},\mathbf{A}^{-1}]
\end{equation}
Therefore, we can invert the matrix $\mathbf{A}$ simply by collecting voltage samples at thermal equilibrium and computing the sample covariance matrix. (This assumes that $\mathbf{A}$ is a positive semi-definite (PSD) matrix, although the extension to non-PSD matrices is possible with a pre-processing step~\cite{aifer2023thermodynamic}.)

Figure~\ref{fig:Inversion4x4} shows the results of inverting a 4x4 matrix on the SPU, which involves using only a 4-dimensional subset of the unit cells on the SPU. One can see the error (i.e., the relative Frobenius error between the SPU inverse and the target inverse) go down as the number of samples increases. The SPU inverse after gathering several thousand samples looks visually similar to the target inverse. The fact that the error does not go exactly to zero indicates that experimental imperfections are present. This likely includes capacitance tolerances, i.e., nominal capacitance values being off from their true values.

\begin{figure}[t]%
\centering
\includegraphics[width=0.55\textwidth]{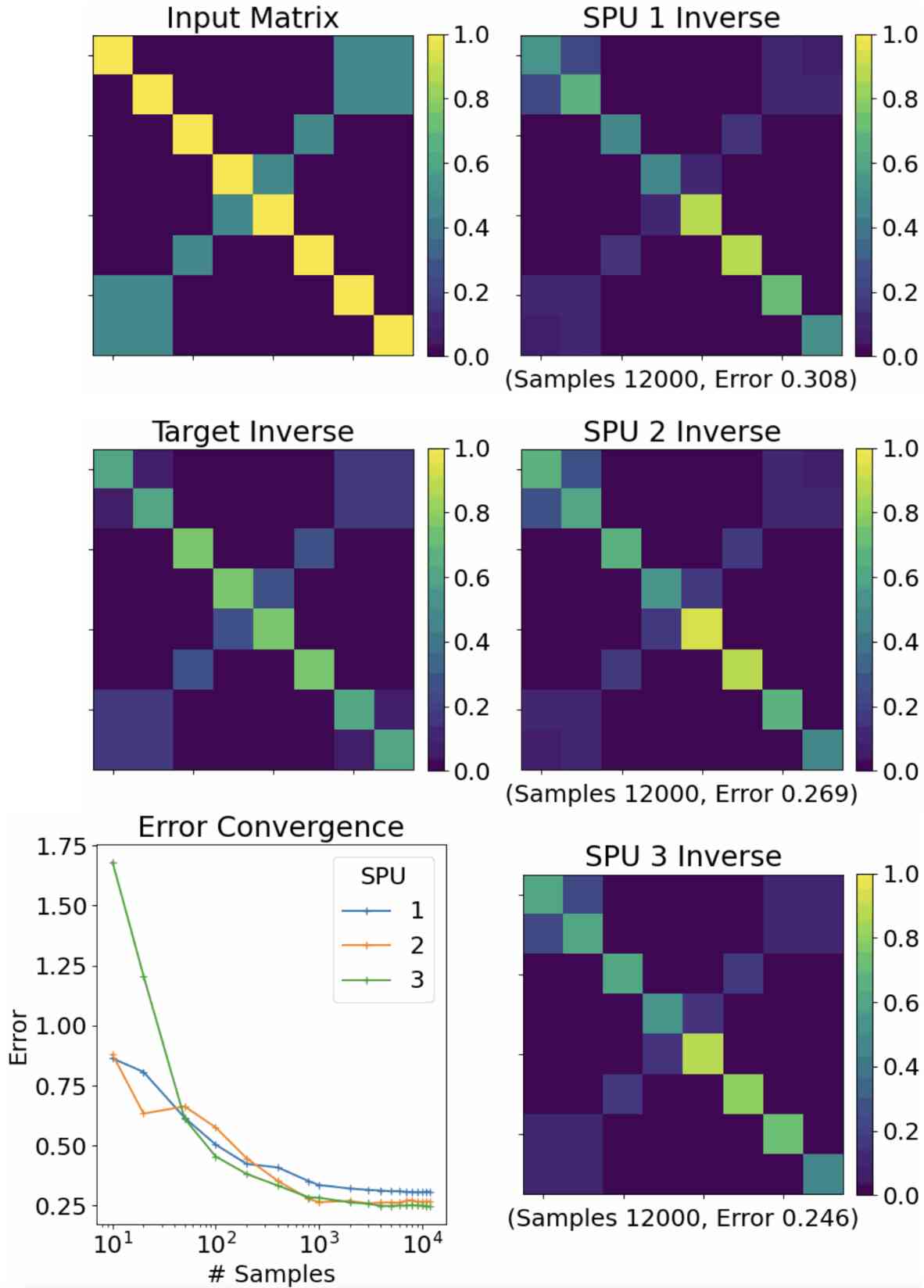}
\caption{\justifying \textbf{Thermodynamic inversion of an 8x8 matrix.} This experiment was performed independently on three distinct (but nominally identical) copies of the SPU. The input matrix $A$ and its true inverse $A^{-1}$ are shown, respectively, on the top left and middle left. The relative Frobenius error versus the number of samples is plotted in the bottom left, for each of the three SPUs. The three panels on the right show the experimentally determined inverses after gathering 12000 samples on each of three SPUs.}\label{fig:Inversion8x8}
\end{figure}

\begin{figure}[h]%
\centering
\includegraphics[width=0.995\textwidth]{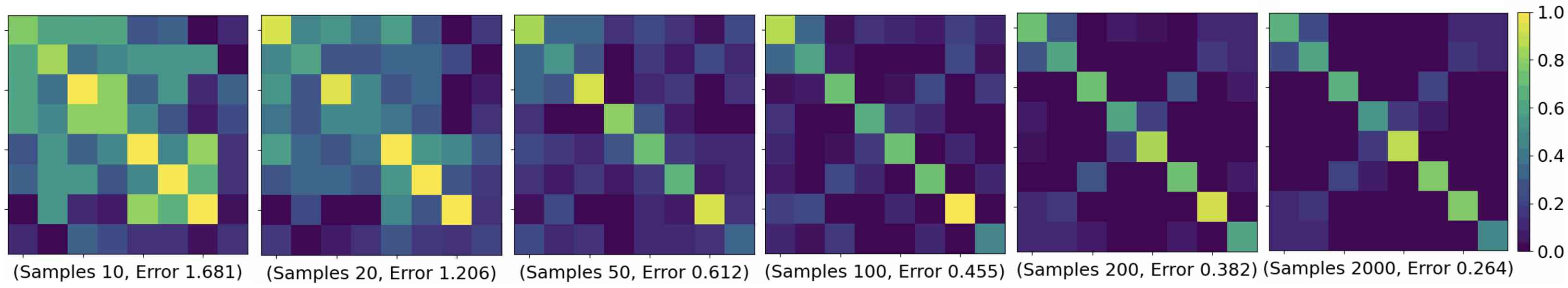}
\caption{\justifying\textbf{Time progression of thermodynamic matrix inversion.} From left to right, more samples are gathered from the SPU to compute the matrix inversion. The number of samples and the inversion error are stated below each panel. One can visually see the resulting inverse look more like the target inverse (shown in Fig.~\ref{fig:Inversion8x8}) as more samples are obtained.}\label{fig:InversionVsSamples}
\end{figure}

Figure~\ref{fig:Inversion8x8} shows the 8x8 inversion results. In this case, we perform the algorithm on three distinct copies of the SPU, which are nominally identical (although may slightly differ due to component tolerances). The fact that similar results are obtained on all three SPUs (as shown in Fig.~\ref{fig:Inversion8x8}) is useful for demonstrating scientific reproducibility. Indeed, one can see the relative Frobenius error go down as the number of samples increases. (Once again the error does not go exactly to zero due to experimental imperfections.) The final SPU inverse looks visually similar to target inverse. Moreover, Fig.~\ref{fig:InversionVsSamples} shows the time evolution of the SPU inverse. One can see in Fig.~\ref{fig:InversionVsSamples} that the SPU inverse gradually looks more-and-more like the true inverse as more samples are gathered.

In the Supplemental Information, we show some additional plots for matrix inversion on our SPU, including the dependence of the error on the condition number of $A$ and on the smallest eigenvalue of $A$.

\subsection{Gaussian Process Regression} 

We now illustrate Gaussian process regression (GPR) with our SPU. GPR is a key primitive in probabilistic machine learning~\cite{murphy2022probabilistic} and  essentially corresponds to performing regression with uncertainty. We consider one-dimensional data  $\mathcal{D} = \{(x_i, y_i)\}$ where $i \in \{1,...,n_{train}\}$ and we want to make estimates about the values $\tilde{y}_j$ at points $\tilde{x}_j$ where $j \in \{1,...,n_{test}\}$.

The underlying assumption of GPR is that both $y$ and $\tilde{y}$ are elements of a vector that is produced by sampling a multivariate Gaussian distribution with some mean and covariance matrix. Therefore, we can split this distribution and write:
\begin{equation}
\begin{pmatrix}
y \\
\tilde{y}
\end{pmatrix}
= \mathcal{N}\bigg[
\begin{pmatrix}
\mu \\
\tilde{\mu}
\end{pmatrix},
\begin{pmatrix}
\Sigma_{11} & \Sigma_{12} \\
\Sigma_{12} & \Sigma_{22}
\end{pmatrix}\bigg]
\end{equation}
where $\Sigma_{11} = k(x, x)$, $\Sigma_{12} = k(x, \tilde{x})$, $\Sigma_{22} = k(\tilde{x},\tilde{x})$, and $k(\cdot,\cdot)$ denotes the kernel function.  So $\Sigma_{11}$ has dimensions $n_{train} \times n_{train}$, $\Sigma_{12}$ dimensions $n_{train} \times n_{test}$, and $\Sigma_{22}$ dimensions $n_{test} \times n_{test}$. 

The goal in GPR is to determine the posterior distribution at a given test point, which can be written as $P(\tilde{y}|x, \tilde{x}, y) = \mathcal{N}(\mu_{2|1}, \Sigma_{2|1})$. Using properties of Gaussians one can show that:
\begin{equation}
\mu_{2|1} = \Sigma_{21}\Sigma_{11}^{-1}y,\qquad \Sigma_{2|1} = \Sigma_{22} - \Sigma_{21} \Sigma_{11}^{-1}\Sigma_{12}
\end{equation}
In order to calculate the mean and the covariance of the posterior distribution at the test data points we need to invert $\Sigma_{11}$, which is an $n_{train} \times n_{train}$ matrix. Therefore, the algorithmic complexity is dominated by the matrix inversion, that is $O(n_{train}^{3})$. Because this is the dominant computational step, in our experimental implementation we utilize our SPU for the matrix inversion subroutine and use a digital device for the other subroutines of GPR.

\begin{figure}[t]
    \centering
    \includegraphics[scale=0.6]{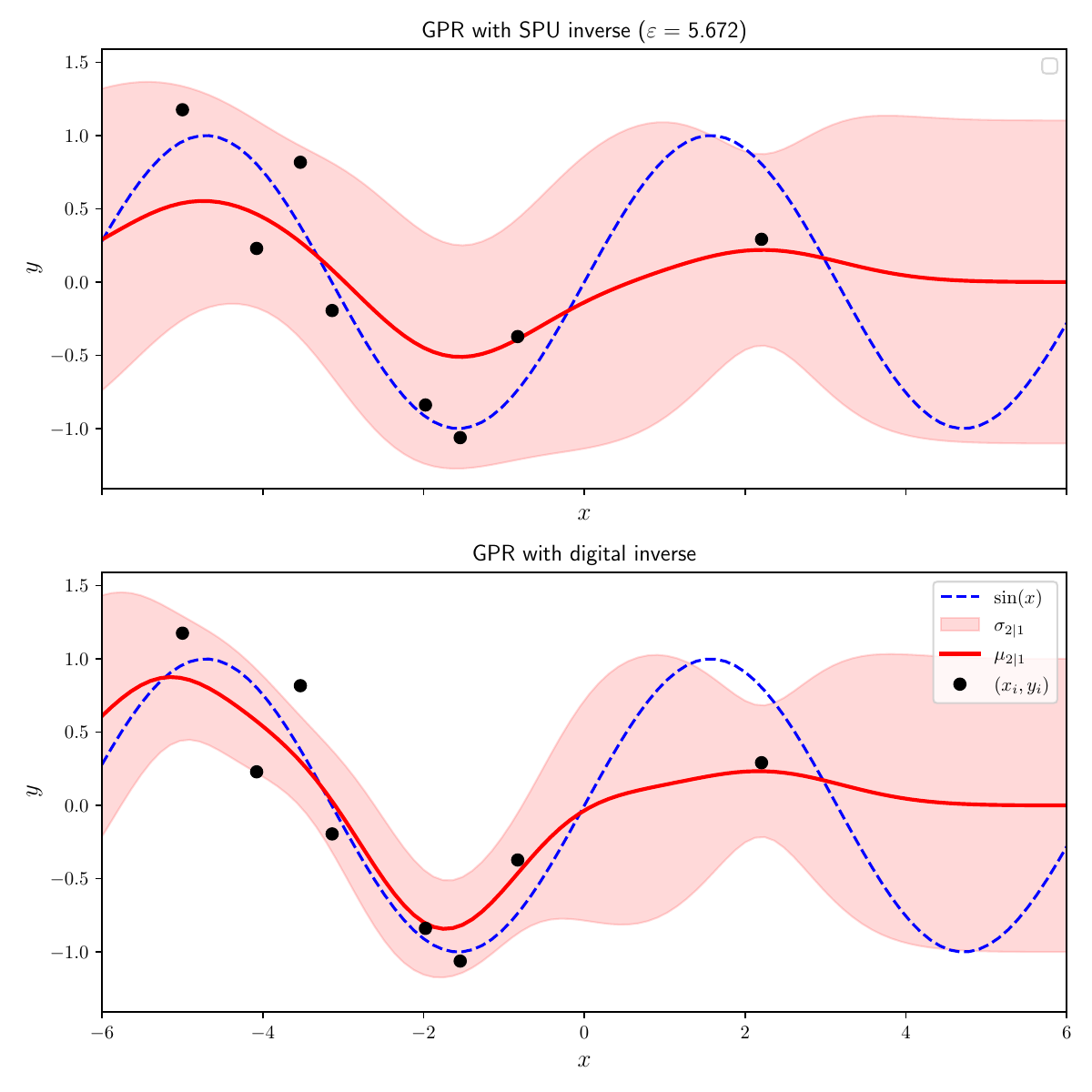}
    \caption{\justifying\textbf{GPR results using the stochastic processing unit.} The black dots represent noisy training data, generated from adding noise to a sine function (dashed blue line). The red line represents the posterior mean prediction $\mu_{2|1}$ and the light-red region represents the posterior variances at each point $\sigma_{2|1}$. The top (bottom) panel show the results from using our SPU (a digital device) for the GPR protocol.}
    \label{fig:gpr_pcb}
\end{figure}

For our experimental implementation, we consider synthetic data generated as $y_i = \sin(x_i) + \epsilon$, with $\epsilon$ being random Gaussian noise of unit variance.  We use 8 training data points, which corresponds to inverting an 8x8 matrix in the GPR protocol (and hence this matrix inversion can fit on our SPU). We use a radial-basis function (RBF) kernel, common for one-dimensional problems. The results are shown in Fig.~\ref{fig:gpr_pcb}, with the top panel corresponding to using our SPU, and the bottom panel showing for comparison the case where a digital device is used for GPR. One can see from the top panel that the inferred mean (red curve) has oscillatory character and is starting to look somewhat sinusoidal. Moreover, the true function $\sin(x)$ largely lies within the error bars (i.e., the inferred standard deviation), suggesting that the GPR protocol is working reasonably well with our SPU. 

\subsection{Uncertainty quantification for neural networks} 

Neural networks enable classification of high-dimensional data and largely provide the foundation for modern machine learning. However, a key shortcoming is the ``overconfidence'' of the predictions made by neural networks. This phenomenon corresponds to out-of-distribution data being classified with high certainty as belonging to one of the learned classes.  Uncertainty quantification (UQ) provides means to tackle the overconfidence issue, and is crucial for making AI more reliable. 

Here we focus on a state-of-the-art method for UQ in neural network classification called spectral-normalized neural Gaussian processes (SNGP)~\cite{liu2020simple}. SNGP involves replacing the output layer of a deep neural network with a Gaussian process, and is more scalable than other standard approaches to UQ while maintaining high quality.

\begin{figure}[t]
        \centering
        \includegraphics[width=0.8\textwidth]{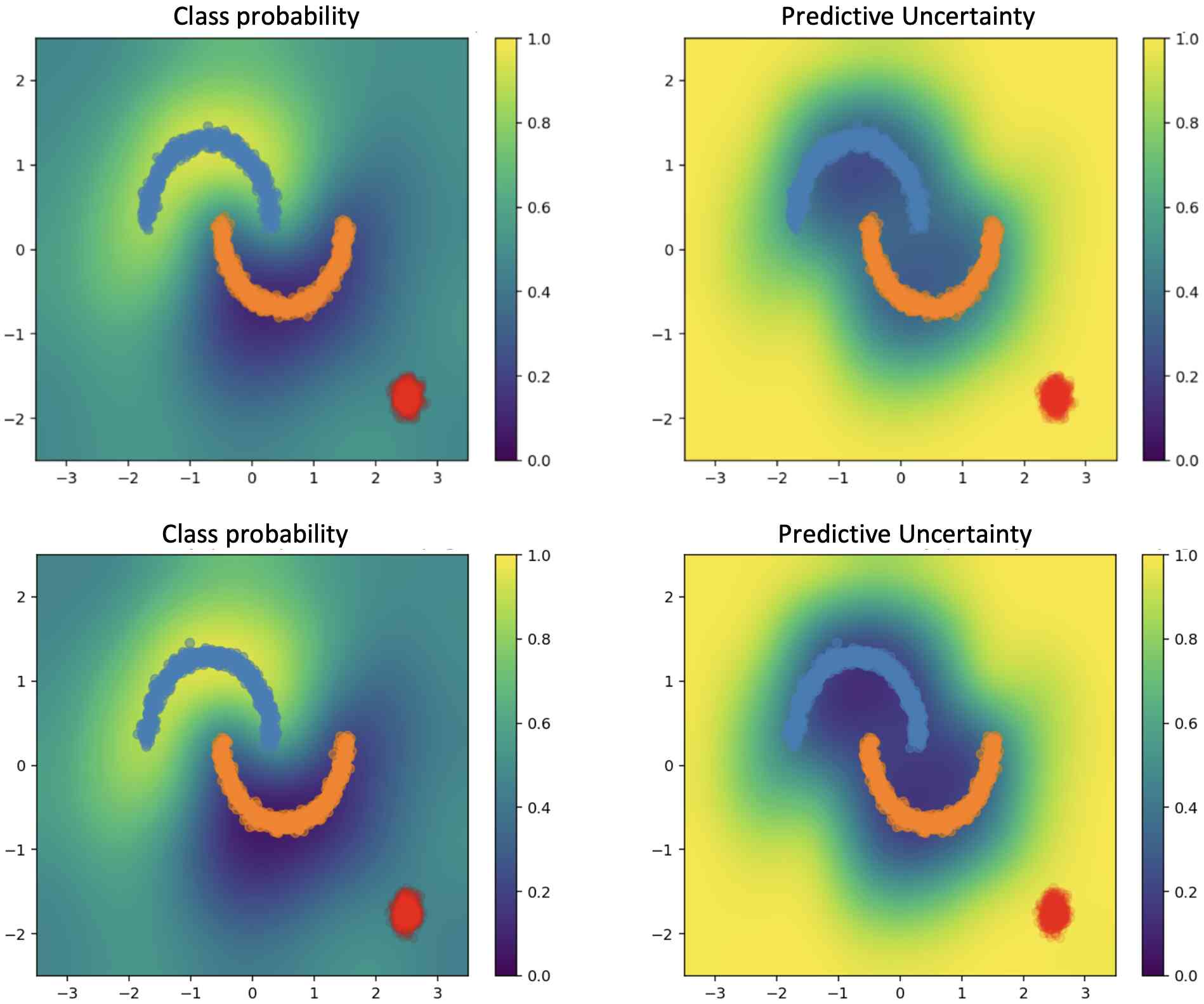}
    \caption{\justifying \textbf{Uncertainty quantification on the output of a ResNet neural network using the SPU.} We consider the two-moons dataset, where the blue and orange points are training points, with color indicating that they belong to the two different classes of the problem, and red points represent out-of-distribution data. The SNGP algorithm is employed to quantify the uncertainty of the predictions. The top and bottom panels respectively correspond to running the SNGP algorithm using a digital device (with Cholesky sampling) and using our SPU to generate samples. The left panels show the class probability, representing the neural network's prediction of points belonging to either class (1 for blue points, and 0 for orange points). The right panels show normalized uncertainty on those predictions, where 1 represents high uncertainty and 0 a low predictive uncertainty.}
     \label{fig:sngp}
\end{figure}

When estimating uncertainty for class predictions with SNGP, one needs to sample from a multivariate Gaussian, which can become a bottleneck for large test sets. As a proof-of-principle, we show that we can use our SPU to sample from the target multivariate Gaussian and generate uncertainty estimates for out-of-distribution data. We consider the two-moons dataset, shown in Fig.~\ref{fig:sngp}, with the dataset defined as $\mathcal{D} = \{x_i, y_i\}_{i=1}^{N}$, $x_i \in \mathbb{R}^2$ and $y_i \in \{0, 1\}$. This dataset consists of two sets of points that are not linearly separable (each point belonging either to the top or the bottom moon), and is a common benchmarking dataset used for UQ. We use a similar ResNet architecture to ref.~\cite{liu2020simple}, where we do not perform a mean-field approximation on covariance matrix of the target distribution

Figure~\ref{fig:sngp} shows that we get relatively good agreement between performing SNGP on a digital device with Cholesky sampling (top panels) versus using our SPU for sampling (bottom panels). A key feature of the SNGP algorithm is that it yields high (low) uncertainty for test data points that are far away from (close to) the training data. This feature is clearly evident in Fig.~\ref{fig:sngp}, suggesting that our SPU is capable of performing the SNGP algorithm. In the figure, 64 grid points are used, and the grid was split into patches of size 8, for which we could use the 8-dimensional SPU to sample from the Gaussian distributed with zero mean and the target covariance matrix. The mean was then added digitally to the gathered samples.

\subsection{Performance advantage at scale} 

We now focus on the potential advantage that one could obtain from scaling up our SPU. Specifically, we compare the expected runtime and energy consumption of our SPU to that of state-of-the-art GPUs, for the task of Gaussian sampling. Our mathematical model for runtime and energy consumption involves considering the effect of three key stages: calculating the elements of the capacitor array from the input covariance matrix, loading/readout of data, and the integration time of the physical dynamics needed to generate the samples. We assume the SPU is constructed from standard electrical components operating at room temperature, we assume the ideal case of electrical components with 16 bits of precision, and we assume the unit cells in the SPU are fully connected (This analysis is valid whether the user provides the covariance matrix or the precision matrix, where details of sampling when the covariance matrix is provided is discussed in the Supplemental Information.) In our simulations, the input matrix is compiled to the SPU in a digital pre-processing step, and then the dynamics of the SPU are run for the time needed to generate 10,000 samples. We assume a physical time constant of $1\mu$s and that the number of analog-to-digital conversion channels scales with dimension with a sampling rate of $10$ Megasamples per second. This leads to a power consumption estimate of 0.005 mW per cell, assuming it is dominated by the ADC power consumption. For comparison with digital state-of-the-art, we obtain digital timing and energy consumption results using a JAX implementation of Cholesky sampling on an NVIDIA RTX A6000 GPU.

\begin{figure}[t]
    \centering
    \includegraphics[width=0.8\textwidth]{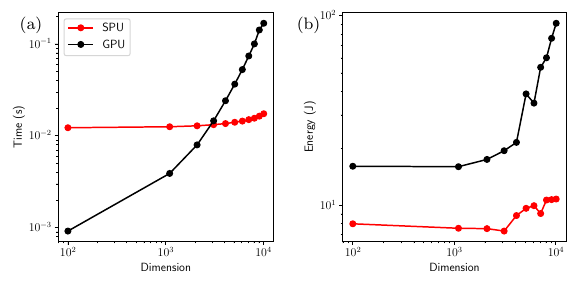}
    \caption{\justifying \textbf{Comparison of GPU and estimated SPU performance.} Panel (a): time to solution to obtain 10,000 Gaussian samples for dimensions ranging from 100 to 10,000 for an A6000 GPU (black dots) and an SPU (red dots). Panel (b): Corresponding energy to solution.}
    \label{fig:time_and_energy_adv}
\end{figure}

Figure~\ref{fig:time_and_energy_adv}(a) shows how the time taken to produce samples from a multivariate Gaussian scales with dimension. One can see that the GPU is faster for low dimensions, but the SPU performance is expected to outperform the GPU for high dimensions. The cross-over point, which we call the point of ``thermodynamic advantage'', occurs around $d=3000$ for the assumptions we made. The asymptotic scaling for high dimensions is expected to go as $O(d^2)$ for the SPU, as opposed to $O(d^3)$ for Cholesky sampling on the GPU. At $d=10,000$, an order-of-magnitude speedup is predicted, and larger speedups could be unlocked by scaling up even larger.

It is also of interest to consider energy, since it is well known that GPUs consume large amounts of energy. In contrast, the natural dynamics of a physical system are energy frugal and could reduce the expected energy requirements. Figure~\ref{fig:time_and_energy_adv}(b) shows how the energy of generating Gaussian samples is expected to scale with dimension for both the SPU and GPU. One can see that our model predicts the exciting prospect that the SPU provides energy savings for all dimensions, even for low dimensions. Moreover, the energy saving is an order-of-magnitude at $d=10,000$ and is expected to continue to grow for larger dimensions.

Ultimately, these results represent a mathematical model, and the true evidence of thermodynamic advantage will only be obtained by directly scaling up the SPU hardware. Nevertheless, it is encouraging that a simple model of the SPU, the timings involved in its end-to-end operation, and the energy cost during these processes lead to a potential speedup and energy savings of more than an order-of-magnitude relative to state-of-the-art GPUs with relatively conservative assumptions. While we showed results here for the case of Gaussian sampling, we note that similar potential speedups and energy advantages for matrix inversion are expected.

\section{Discussion}\label{sec12}

The field of thermodynamic computing is in its early days. This is similar to quantum computing in the 1990's, when Shor discovered quantum error correction~\cite{shor1995scheme} and his famous quantum factoring algorithm~\cite{shor1999polynomial}, and Chuang, Gershenfeld, and Kubinec built the first small-scale quantum computer~\cite{chuang1998experimental}. If the field thermodynamic computing follows the same trajectory as that of quantum computing, then one can expect thermodynamic computing to permeate academic research labs in the near future, and the work described herein would represent a historical landmark for the field, as the beginning of a new paradigm.  

In building the first CV thermodynamic computer, we provided a blueprint for how to perform thermodynamic computations for Gaussian sampling and linear algebra, as well as applications that utilize these primitives like regression. Indeed our matrix inversion experiments were the first implementations of thermodynamic linear algebra, hopefully inspiring further investigations into this exciting new sub-field of thermodynamic computing. Moreover, our work provides the first step towards achieving thermodynamic advantage, which corresponds to building a large enough thermodynamic computer so as to outperform state-of-the-art digital computers in either speed or energy efficiency. Achieving this would represent a huge breakthrough for the field of AI, which already consumes vast computing and energy resources.

As our hardware is particularly relevant to probabilistic AI, we note that probabilistic reasoning is central to unlocking reliable AI for high-stakes enterprise applications, where risk has been a central barrier to AI adoption. Moreover, it may even be central to unlocking artificial general intelligence (AGI). This follows from recent arguments~\cite{LeCun2023MITphysics} that current AI technology, which is largely based on transformers, is not capable of reasoning and planning and hence not capable, on its own, of achieving AGI. Our computing hardware, when appropriately scaled up, will enable and accelerate probabilistic reasoning for AI applications. As an example, we demonstrated how our hardware enables uncertainty quantification on the outputs of neural networks. We, therefore, envision a key role for our hardware paradigm in bringing reliable AI to enterprise applications, and even in bringing reasoning capabilities to AGI efforts.

\section{Methods}\label{sec11}

Here we provide additional details on the design of our hardware, including the interface to digital hardware, the coupled resonator structure, the noise source, and the readout process.

\subsection{Interface to digital hardware} 

\begin{figure}[t]%
\centering
\includegraphics[width=0.9\textwidth]{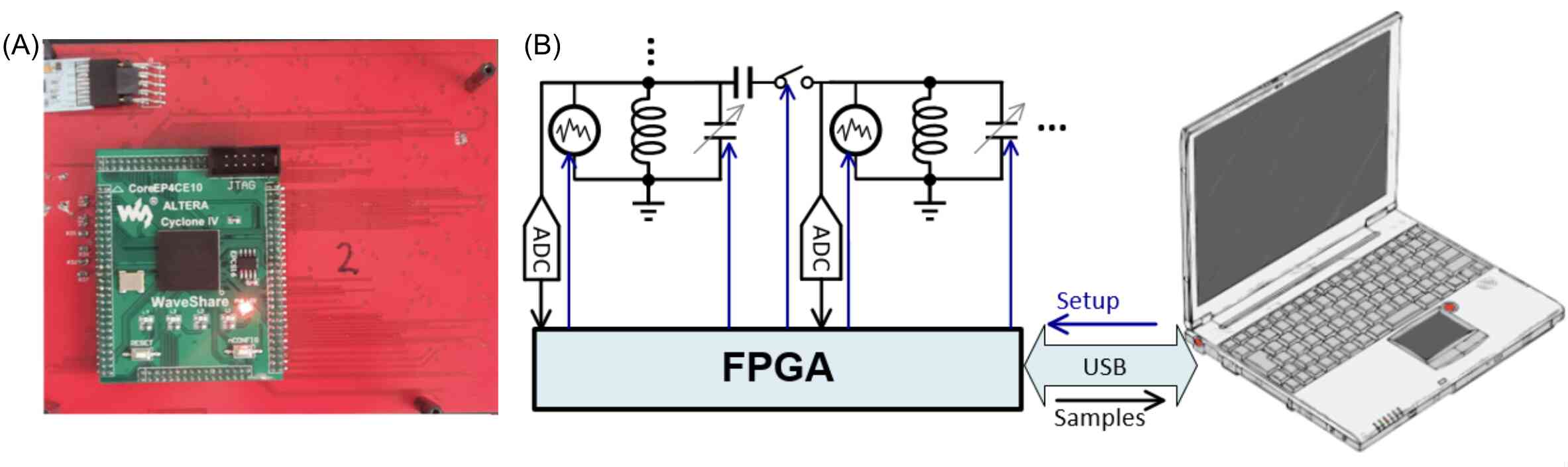}
\caption{\justifying \textbf{Interface of SPU with Digital Hardware.} The SPU is controlled by a digital device as a co-processor. Panel (A) shows the back of the circuit board, displaying both the FPGA and the USB that leads to the CPU. Panel (B) shows schematically how the data flows between the analog and the digital devices. The data that determines the circuit parameters is downloaded from the CPU to the SPU, and the sampling data is uploaded from the SPU to the CPU. The FPGA acts as an intermediary in this process, and also provides the noise source.}\label{fig:system_level}
\end{figure}

Figure~\ref{fig:system_level} shows the interface between the analog and digital hardware for our thermodynamic computing system. Broadly speaking, our SPU functions as a co-processor to a digital device. Namely, the operation of the SPU is controlled by a Central Processing Unit (CPU) and a Field-Programmable Gate Array (FPGA). 

Figure~\ref{fig:system_level}(A) shows the back of the circuit board for the SPU. One can see the FPGA attached to the board as well as the Universal Serial Bus (USB) that leads to the CPU. The CPU compiles the requested covariance matrix into FPGA code, while the FPGA opens and closes the switches controlling the capacitance values and the coupling branches and also starts the noise sources, as shown in Fig.~\ref{fig:system_level}(B). The FPGA then starts the measurement phase where voltage measurements are taken at a given rate, then digitized by an onboard analog-to-digital converter (ADC) and sent to the CPU.

\subsection{Coupled resonators structure}\label{subsec2}

\begin{figure}[h]%
\centering
\includegraphics[width=0.9\textwidth]{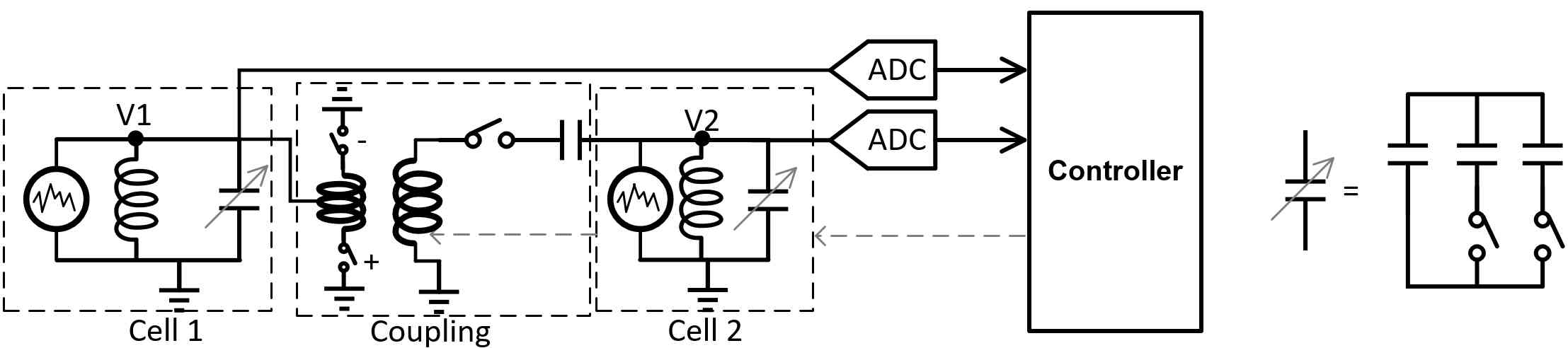}
\caption{\justifying \textbf{A simplified version of the SPU, containing two cells with a single coupling branch.} In-cell capacitances arise from a bank of three capacitors as shown. The coupling capacitance arises from a single switched capacitor. A transformer appears in the coupling circuit, for the purpose of achieving bipolar coupling.}\label{fig_2cell}
\end{figure}

While Fig.~\ref{fig:PCB_AllAll} shows a high-level view of the circuit structure, we give a more precise depiction in Fig.~\ref{fig_2cell}. (For simplicity, we only show two unit cells in Fig.~\ref{fig_2cell}, even though our SPU has 8 cells.) Each cell consists of an inductor, a tunable capacitor, and a current noise source that is uncorrelated to other noise sources in the SPU. While ideally a continuously tunable capacitor is desired, available technologies (e.g., varactor diodes or BST-based varactors) typically suffer from nonlinearity or complexity of integration. Consequently, we use switched capacitors in a capacitor bank. Simulations show that, with as little as three capacitors, the quality of operation is nearly unaffected. The exact values of the capacitors were chosen with numerical optimization to obtain the best performance.

The coupling between the resonators is implemented capacitively. Similar to the capacitors in the cells, the coupling capacitors should ideally be continuously controlled, but a switched capacitor is used due to the limitations of tunable capacitors. A center-tapped transformer is used to switch between positive and negative polarity coupling, which allows one to achieve bipolar coupling. 

The nominal capacitances for the in-cell capacitors are chosen from the following set of values: 
$$\text{In-cell capacitances}\in \{1.0 \text{nF}, 3.2 \text{nF}, 4.3 \text{nF}, 6.5 \text{nF}\}.$$ 
The nominal coupling capacitance is 0.47 nF through a center-tapped transformer, making the available coupling values: 
$$\text{Coupling values}\in\{ -x, 0, x \}.$$ 
where $x=0.47 \times 10^{-9}$.

\subsection{Noise source}\label{subsec_pn}

The noise is generated using Linear-feedback shift register (LFSR) from the FPGA. An RC lowpass filter (LPF) is used to suppress the high frequency components of the digital output of the FPGA. In order to minimize the loading effect of the digital source on each cell (which would reduce the quality factor), a large resistance is used in series with the digital noise pin (Thevenin's equivalent of the current source). 

\begin{figure}[h]%
\centering
\includegraphics[width=0.9\textwidth]{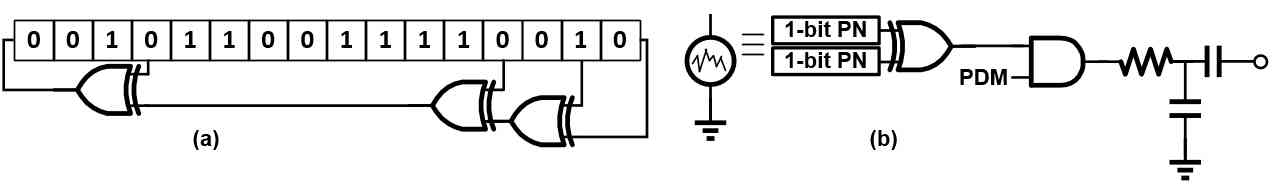}
\caption{\justifying (a) Noise source for each cell. (b) 16-bit LFSR pseudo-random bit sequence generator.}\label{fig_lfsr}
\end{figure}

The pseudo-noise sequences are based on a 16-bit primitive polynomial for maximum sequence length, shown in Figure~\ref{fig_lfsr}(a). To ensure that the sequences in each cell are uncorrelated to the ones in other cells, each LFSR starts with a different initial condition. The primitive polynomial is given by:
\begin{equation}
f(x)=x^{16}+x^{15}+x^{13}+x^{4}+1.\label{eq_pn}
\end{equation}

An LFSR as the one shown in Fig.~\ref{fig_lfsr}(a) produces a uniform distribution. To bring this a step closer to a Gaussian distribution, we add the outputs of two LFSRs (using an XOR), delivering a triangular distribution. The resulting bit sequence is typically called gold codes. The RC filter at the output, shown in Figure \ref{fig_lfsr}(b), smooths that distribution, which approximates a Gaussian distribution.

For proper operation of the SPU, the noise variance has to be set to a given value. The output of the LFSR, however, has a constant RMS value. This can be resolved by duty-cycling the output of the LFSR. We use Pulse Density Modulation (PDM) with an AND gate as shown in Figure \ref{fig_lfsr}(b).

\subsection{Readout of samples}\label{subsec2_adc}

Reading the voltage across each cell occurs after the capacitance values are set, and the noise source is initiated. In order to collect sufficiently large number of samples to be statistically significant, the sampling rate is expected to be relatively high. On the other hand, high sampling rate might result in non-zero correlation between a sample and a subsequent one. The time until samples become uncorrelated is inversely proportional to the quality factor ($Q$) of the LC resonators. For this implementation, the optimal sampling rate was found to be 12 MHz, using an eight-channel 10-bit ADS5292 ADC.

\bibliography{thermo.bib}

\begin{thebibliography}{31}%
\makeatletter
\providecommand \@ifxundefined [1]{%
 \@ifx{#1\undefined}
}%
\providecommand \@ifnum [1]{%
 \ifnum #1\expandafter \@firstoftwo
 \else \expandafter \@secondoftwo
 \fi
}%
\providecommand \@ifx [1]{%
 \ifx #1\expandafter \@firstoftwo
 \else \expandafter \@secondoftwo
 \fi
}%
\providecommand \natexlab [1]{#1}%
\providecommand \enquote  [1]{``#1''}%
\providecommand \bibnamefont  [1]{#1}%
\providecommand \bibfnamefont [1]{#1}%
\providecommand \citenamefont [1]{#1}%
\providecommand \href@noop [0]{\@secondoftwo}%
\providecommand \href [0]{\begingroup \@sanitize@url \@href}%
\providecommand \@href[1]{\@@startlink{#1}\@@href}%
\providecommand \@@href[1]{\endgroup#1\@@endlink}%
\providecommand \@sanitize@url [0]{\catcode `\\12\catcode `\$12\catcode `\&12\catcode `\#12\catcode `\^12\catcode `\_12\catcode `\%12\relax}%
\providecommand \@@startlink[1]{}%
\providecommand \@@endlink[0]{}%
\providecommand \url  [0]{\begingroup\@sanitize@url \@url }%
\providecommand \@url [1]{\endgroup\@href {#1}{\urlprefix }}%
\providecommand \urlprefix  [0]{URL }%
\providecommand \Eprint [0]{\href }%
\providecommand \doibase [0]{http://dx.doi.org/}%
\providecommand \selectlanguage [0]{\@gobble}%
\providecommand \bibinfo  [0]{\@secondoftwo}%
\providecommand \bibfield  [0]{\@secondoftwo}%
\providecommand \translation [1]{[#1]}%
\providecommand \BibitemOpen [0]{}%
\providecommand \bibitemStop [0]{}%
\providecommand \bibitemNoStop [0]{.\EOS\space}%
\providecommand \EOS [0]{\spacefactor3000\relax}%
\providecommand \BibitemShut  [1]{\csname bibitem#1\endcsname}%
\let\auto@bib@innerbib\@empty
\bibitem [{\citenamefont {Hooker}(2021)}]{hooker2021hardware}%
  \BibitemOpen
  \bibfield  {author} {\bibinfo {author} {\bibfnamefont {Sara}\ \bibnamefont {Hooker}},\ }\bibfield  {title} {\enquote {\bibinfo {title} {The hardware lottery},}\ }\href@noop {} {\bibfield  {journal} {\bibinfo  {journal} {Communications of the ACM}\ }\textbf {\bibinfo {volume} {64}},\ \bibinfo {pages} {58--65} (\bibinfo {year} {2021})}\BibitemShut {NoStop}%
\bibitem [{\citenamefont {Izmailov}\ \emph {et~al.}(2021)\citenamefont {Izmailov}, \citenamefont {Vikram}, \citenamefont {Hoffman},\ and\ \citenamefont {Wilson}}]{izmailov2021bayesian}%
  \BibitemOpen
  \bibfield  {author} {\bibinfo {author} {\bibfnamefont {Pavel}\ \bibnamefont {Izmailov}}, \bibinfo {author} {\bibfnamefont {Sharad}\ \bibnamefont {Vikram}}, \bibinfo {author} {\bibfnamefont {Matthew~D}\ \bibnamefont {Hoffman}}, \ and\ \bibinfo {author} {\bibfnamefont {Andrew Gordon~Gordon}\ \bibnamefont {Wilson}},\ }\bibfield  {title} {\enquote {\bibinfo {title} {What are bayesian neural network posteriors really like?}}\ }in\ \href@noop {} {\emph {\bibinfo {booktitle} {International conference on machine learning}}}\ (\bibinfo {organization} {PMLR},\ \bibinfo {year} {2021})\ pp.\ \bibinfo {pages} {4629--4640}\BibitemShut {NoStop}%
\bibitem [{\citenamefont {LeCun}(2023)}]{LeCun2023MITphysics}%
  \BibitemOpen
  \bibfield  {author} {\bibinfo {author} {\bibfnamefont {Yann}\ \bibnamefont {LeCun}},\ }\href@noop {} {\enquote {\bibinfo {title} {Objective-driven {AI}},}\ }\bibinfo {howpublished} {\url{https://www.youtube.com/watch?v=vyqXLJsmsrk}} (\bibinfo {year} {2023})\BibitemShut {NoStop}%
\bibitem [{\citenamefont {Hinton}(2022)}]{HintonNeurIPS2022}%
  \BibitemOpen
  \bibfield  {author} {\bibinfo {author} {\bibfnamefont {Geoffrey}\ \bibnamefont {Hinton}},\ }\href@noop {} {\enquote {\bibinfo {title} {Neurips 2022},}\ }\bibinfo {howpublished} {\url{https://neurips.cc/Conferences/2022/ScheduleMultitrack?event=55869}} (\bibinfo {year} {2022})\BibitemShut {NoStop}%
\bibitem [{\citenamefont {Mohseni}\ \emph {et~al.}(2022)\citenamefont {Mohseni}, \citenamefont {McMahon},\ and\ \citenamefont {Byrnes}}]{mohseni2022ising}%
  \BibitemOpen
  \bibfield  {author} {\bibinfo {author} {\bibfnamefont {Naeimeh}\ \bibnamefont {Mohseni}}, \bibinfo {author} {\bibfnamefont {Peter~L.}\ \bibnamefont {McMahon}}, \ and\ \bibinfo {author} {\bibfnamefont {Tim}\ \bibnamefont {Byrnes}},\ }\bibfield  {title} {\enquote {\bibinfo {title} {Ising machines as hardware solvers of combinatorial optimization problems},}\ }\href {\doibase 10.1038/s42254-022-00440-8} {\bibfield  {journal} {\bibinfo  {journal} {Nat. Rev. Phys.}\ }\textbf {\bibinfo {volume} {4}},\ \bibinfo {pages} {363--379} (\bibinfo {year} {2022})}\BibitemShut {NoStop}%
\bibitem [{\citenamefont {Aadit}\ \emph {et~al.}(2022)\citenamefont {Aadit}, \citenamefont {Grimaldi}, \citenamefont {Carpentieri}, \citenamefont {Theogarajan}, \citenamefont {Martinis}, \citenamefont {Finocchio},\ and\ \citenamefont {Camsari}}]{aadit2022massively}%
  \BibitemOpen
  \bibfield  {author} {\bibinfo {author} {\bibfnamefont {Navid~Anjum}\ \bibnamefont {Aadit}}, \bibinfo {author} {\bibfnamefont {Andrea}\ \bibnamefont {Grimaldi}}, \bibinfo {author} {\bibfnamefont {Mario}\ \bibnamefont {Carpentieri}}, \bibinfo {author} {\bibfnamefont {Luke}\ \bibnamefont {Theogarajan}}, \bibinfo {author} {\bibfnamefont {John~M.}\ \bibnamefont {Martinis}}, \bibinfo {author} {\bibfnamefont {Giovanni}\ \bibnamefont {Finocchio}}, \ and\ \bibinfo {author} {\bibfnamefont {Kerem~Y.}\ \bibnamefont {Camsari}},\ }\bibfield  {title} {\enquote {\bibinfo {title} {Massively parallel probabilistic computing with sparse {I}sing machines},}\ }\href {\doibase 10.1038/s41928-022-00774-2} {\bibfield  {journal} {\bibinfo  {journal} {Nat. Electron.}\ }\textbf {\bibinfo {volume} {5}},\ \bibinfo {pages} {460--468} (\bibinfo {year} {2022})}\BibitemShut {NoStop}%
\bibitem [{\citenamefont {Mourgias-Alexandris}\ \emph {et~al.}(2023)\citenamefont {Mourgias-Alexandris}, \citenamefont {Ballani}, \citenamefont {Berloff}, \citenamefont {Clegg}, \citenamefont {Cletheroe}, \citenamefont {Gkantsidis}, \citenamefont {Haller}, \citenamefont {Lyutsarev}, \citenamefont {Parmigiani}, \citenamefont {Pickup} \emph {et~al.}}]{mourgias2023analog}%
  \BibitemOpen
  \bibfield  {author} {\bibinfo {author} {\bibfnamefont {George}\ \bibnamefont {Mourgias-Alexandris}}, \bibinfo {author} {\bibfnamefont {Hitesh}\ \bibnamefont {Ballani}}, \bibinfo {author} {\bibfnamefont {Natalia~G}\ \bibnamefont {Berloff}}, \bibinfo {author} {\bibfnamefont {James~H}\ \bibnamefont {Clegg}}, \bibinfo {author} {\bibfnamefont {Daniel}\ \bibnamefont {Cletheroe}}, \bibinfo {author} {\bibfnamefont {Christos}\ \bibnamefont {Gkantsidis}}, \bibinfo {author} {\bibfnamefont {Istvan}\ \bibnamefont {Haller}}, \bibinfo {author} {\bibfnamefont {Vassily}\ \bibnamefont {Lyutsarev}}, \bibinfo {author} {\bibfnamefont {Francesca}\ \bibnamefont {Parmigiani}}, \bibinfo {author} {\bibfnamefont {Lucinda}\ \bibnamefont {Pickup}},  \emph {et~al.},\ }\bibfield  {title} {\enquote {\bibinfo {title} {Analog iterative machine (aim): using light to solve quadratic optimization problems with mixed variables},}\ }\href@noop {} {\bibfield  {journal} {\bibinfo  {journal} {arXiv preprint arXiv:2304.12594}\ } (\bibinfo {year}
  {2023})}\BibitemShut {NoStop}%
\bibitem [{\citenamefont {Inagaki}\ \emph {et~al.}(2016)\citenamefont {Inagaki}, \citenamefont {Haribara}, \citenamefont {Igarashi}, \citenamefont {Sonobe}, \citenamefont {Tamate}, \citenamefont {Honjo}, \citenamefont {Marandi}, \citenamefont {McMahon}, \citenamefont {Umeki}, \citenamefont {Enbutsu}, \citenamefont {Tadanaga}, \citenamefont {Takenouchi}, \citenamefont {Aihara}, \citenamefont {Kawarabayashi}, \citenamefont {Inoue}, \citenamefont {Utsunomiya},\ and\ \citenamefont {Takesue}}]{inagaki2016coherent}%
  \BibitemOpen
  \bibfield  {author} {\bibinfo {author} {\bibfnamefont {Takahiro}\ \bibnamefont {Inagaki}}, \bibinfo {author} {\bibfnamefont {Yoshitaka}\ \bibnamefont {Haribara}}, \bibinfo {author} {\bibfnamefont {Koji}\ \bibnamefont {Igarashi}}, \bibinfo {author} {\bibfnamefont {Tomohiro}\ \bibnamefont {Sonobe}}, \bibinfo {author} {\bibfnamefont {Shuhei}\ \bibnamefont {Tamate}}, \bibinfo {author} {\bibfnamefont {Toshimori}\ \bibnamefont {Honjo}}, \bibinfo {author} {\bibfnamefont {Alireza}\ \bibnamefont {Marandi}}, \bibinfo {author} {\bibfnamefont {Peter~L.}\ \bibnamefont {McMahon}}, \bibinfo {author} {\bibfnamefont {Takeshi}\ \bibnamefont {Umeki}}, \bibinfo {author} {\bibfnamefont {Koji}\ \bibnamefont {Enbutsu}}, \bibinfo {author} {\bibfnamefont {Osamu}\ \bibnamefont {Tadanaga}}, \bibinfo {author} {\bibfnamefont {Hirokazu}\ \bibnamefont {Takenouchi}}, \bibinfo {author} {\bibfnamefont {Kazuyuki}\ \bibnamefont {Aihara}}, \bibinfo {author} {\bibfnamefont {Ken-ichi}\ \bibnamefont {Kawarabayashi}}, \bibinfo {author}
  {\bibfnamefont {Kyo}\ \bibnamefont {Inoue}}, \bibinfo {author} {\bibfnamefont {Shoko}\ \bibnamefont {Utsunomiya}}, \ and\ \bibinfo {author} {\bibfnamefont {Hiroki}\ \bibnamefont {Takesue}},\ }\bibfield  {title} {\enquote {\bibinfo {title} {A coherent ising machine for 2000-node optimization problems},}\ }\href {\doibase 10.1126/science.aah4243} {\bibfield  {journal} {\bibinfo  {journal} {Science}\ }\textbf {\bibinfo {volume} {354}},\ \bibinfo {pages} {603--606} (\bibinfo {year} {2016})}\BibitemShut {NoStop}%
\bibitem [{\citenamefont {Moy}\ \emph {et~al.}(2022)\citenamefont {Moy}, \citenamefont {Ahmed}, \citenamefont {Chiu}, \citenamefont {Moy}, \citenamefont {Sapatnekar},\ and\ \citenamefont {Kim}}]{moy20221}%
  \BibitemOpen
  \bibfield  {author} {\bibinfo {author} {\bibfnamefont {William}\ \bibnamefont {Moy}}, \bibinfo {author} {\bibfnamefont {Ibrahim}\ \bibnamefont {Ahmed}}, \bibinfo {author} {\bibfnamefont {Po-wei}\ \bibnamefont {Chiu}}, \bibinfo {author} {\bibfnamefont {John}\ \bibnamefont {Moy}}, \bibinfo {author} {\bibfnamefont {Sachin~S}\ \bibnamefont {Sapatnekar}}, \ and\ \bibinfo {author} {\bibfnamefont {Chris~H}\ \bibnamefont {Kim}},\ }\bibfield  {title} {\enquote {\bibinfo {title} {A 1,968-node coupled ring oscillator circuit for combinatorial optimization problem solving},}\ }\href@noop {} {\bibfield  {journal} {\bibinfo  {journal} {Nature Electronics}\ }\textbf {\bibinfo {volume} {5}},\ \bibinfo {pages} {310--317} (\bibinfo {year} {2022})}\BibitemShut {NoStop}%
\bibitem [{\citenamefont {Chou}\ \emph {et~al.}(2019)\citenamefont {Chou}, \citenamefont {Bramhavar}, \citenamefont {Ghosh},\ and\ \citenamefont {Herzog}}]{chou2019analog}%
  \BibitemOpen
  \bibfield  {author} {\bibinfo {author} {\bibfnamefont {Jeffrey}\ \bibnamefont {Chou}}, \bibinfo {author} {\bibfnamefont {Suraj}\ \bibnamefont {Bramhavar}}, \bibinfo {author} {\bibfnamefont {Siddhartha}\ \bibnamefont {Ghosh}}, \ and\ \bibinfo {author} {\bibfnamefont {William}\ \bibnamefont {Herzog}},\ }\bibfield  {title} {\enquote {\bibinfo {title} {Analog coupled oscillator based weighted ising machine},}\ }\href@noop {} {\bibfield  {journal} {\bibinfo  {journal} {Scientific reports}\ }\textbf {\bibinfo {volume} {9}},\ \bibinfo {pages} {14786} (\bibinfo {year} {2019})}\BibitemShut {NoStop}%
\bibitem [{\citenamefont {Wang}\ and\ \citenamefont {Roychowdhury}(2019)}]{wang2019oim}%
  \BibitemOpen
  \bibfield  {author} {\bibinfo {author} {\bibfnamefont {Tianshi}\ \bibnamefont {Wang}}\ and\ \bibinfo {author} {\bibfnamefont {Jaijeet}\ \bibnamefont {Roychowdhury}},\ }\bibfield  {title} {\enquote {\bibinfo {title} {Oim: Oscillator-based ising machines for solving combinatorial optimisation problems},}\ }in\ \href@noop {} {\emph {\bibinfo {booktitle} {Unconventional Computation and Natural Computation: 18th International Conference, UCNC 2019, Tokyo, Japan, June 3--7, 2019, Proceedings 18}}}\ (\bibinfo {organization} {Springer},\ \bibinfo {year} {2019})\ pp.\ \bibinfo {pages} {232--256}\BibitemShut {NoStop}%
\bibitem [{\citenamefont {Theilman}\ \emph {et~al.}(2023)\citenamefont {Theilman}, \citenamefont {Wang}, \citenamefont {Parekh}, \citenamefont {Severa}, \citenamefont {Smith},\ and\ \citenamefont {Aimone}}]{theilman2023stochastic}%
  \BibitemOpen
  \bibfield  {author} {\bibinfo {author} {\bibfnamefont {Bradley~H}\ \bibnamefont {Theilman}}, \bibinfo {author} {\bibfnamefont {Yipu}\ \bibnamefont {Wang}}, \bibinfo {author} {\bibfnamefont {Ojas}\ \bibnamefont {Parekh}}, \bibinfo {author} {\bibfnamefont {William}\ \bibnamefont {Severa}}, \bibinfo {author} {\bibfnamefont {J~Darby}\ \bibnamefont {Smith}}, \ and\ \bibinfo {author} {\bibfnamefont {James~B}\ \bibnamefont {Aimone}},\ }\bibfield  {title} {\enquote {\bibinfo {title} {Stochastic neuromorphic circuits for solving maxcut},}\ }in\ \href@noop {} {\emph {\bibinfo {booktitle} {2023 IEEE International Parallel and Distributed Processing Symposium (IPDPS)}}}\ (\bibinfo {organization} {IEEE},\ \bibinfo {year} {2023})\ pp.\ \bibinfo {pages} {779--787}\BibitemShut {NoStop}%
\bibitem [{\citenamefont {Coles}\ \emph {et~al.}(2023)\citenamefont {Coles}, \citenamefont {Szczepanski}, \citenamefont {Melanson}, \citenamefont {Donatella}, \citenamefont {Martinez},\ and\ \citenamefont {Sbahi}}]{coles2023thermodynamic}%
  \BibitemOpen
  \bibfield  {author} {\bibinfo {author} {\bibfnamefont {Patrick~J.}\ \bibnamefont {Coles}}, \bibinfo {author} {\bibfnamefont {Collin}\ \bibnamefont {Szczepanski}}, \bibinfo {author} {\bibfnamefont {Denis}\ \bibnamefont {Melanson}}, \bibinfo {author} {\bibfnamefont {Kaelan}\ \bibnamefont {Donatella}}, \bibinfo {author} {\bibfnamefont {Antonio~J.}\ \bibnamefont {Martinez}}, \ and\ \bibinfo {author} {\bibfnamefont {Faris}\ \bibnamefont {Sbahi}},\ }\href@noop {} {\enquote {\bibinfo {title} {Thermodynamic {AI} and the fluctuation frontier},}\ } (\bibinfo {year} {2023}),\ \Eprint {http://arxiv.org/abs/2302.06584} {arXiv:2302.06584 [cs.ET]} \BibitemShut {NoStop}%
\bibitem [{\citenamefont {Conte}\ \emph {et~al.}(2019)\citenamefont {Conte}, \citenamefont {DeBenedictis}, \citenamefont {Ganesh}, \citenamefont {Hylton}, \citenamefont {Strachan}, \citenamefont {Williams}, \citenamefont {Alemi}, \citenamefont {Altenberg}, \citenamefont {Crooks}, \citenamefont {Crutchfield} \emph {et~al.}}]{conte2019thermodynamic}%
  \BibitemOpen
  \bibfield  {author} {\bibinfo {author} {\bibfnamefont {Tom}\ \bibnamefont {Conte}}, \bibinfo {author} {\bibfnamefont {Erik}\ \bibnamefont {DeBenedictis}}, \bibinfo {author} {\bibfnamefont {Natesh}\ \bibnamefont {Ganesh}}, \bibinfo {author} {\bibfnamefont {Todd}\ \bibnamefont {Hylton}}, \bibinfo {author} {\bibfnamefont {John~Paul}\ \bibnamefont {Strachan}}, \bibinfo {author} {\bibfnamefont {R~Stanley}\ \bibnamefont {Williams}}, \bibinfo {author} {\bibfnamefont {Alexander}\ \bibnamefont {Alemi}}, \bibinfo {author} {\bibfnamefont {Lee}\ \bibnamefont {Altenberg}}, \bibinfo {author} {\bibfnamefont {Gavin~E.}\ \bibnamefont {Crooks}}, \bibinfo {author} {\bibfnamefont {James}\ \bibnamefont {Crutchfield}},  \emph {et~al.},\ }\bibfield  {title} {\enquote {\bibinfo {title} {Thermodynamic computing},}\ }\href@noop {} {\bibfield  {journal} {\bibinfo  {journal} {arXiv preprint arXiv:1911.01968}\ } (\bibinfo {year} {2019})}\BibitemShut {NoStop}%
\bibitem [{\citenamefont {Aifer}\ \emph {et~al.}(2023)\citenamefont {Aifer}, \citenamefont {Donatella}, \citenamefont {Gordon}, \citenamefont {Ahle}, \citenamefont {Simpson}, \citenamefont {Crooks},\ and\ \citenamefont {Coles}}]{aifer2023thermodynamic}%
  \BibitemOpen
  \bibfield  {author} {\bibinfo {author} {\bibfnamefont {Maxwell}\ \bibnamefont {Aifer}}, \bibinfo {author} {\bibfnamefont {Kaelan}\ \bibnamefont {Donatella}}, \bibinfo {author} {\bibfnamefont {Max~Hunter}\ \bibnamefont {Gordon}}, \bibinfo {author} {\bibfnamefont {Thomas}\ \bibnamefont {Ahle}}, \bibinfo {author} {\bibfnamefont {Daniel}\ \bibnamefont {Simpson}}, \bibinfo {author} {\bibfnamefont {Gavin~E}\ \bibnamefont {Crooks}}, \ and\ \bibinfo {author} {\bibfnamefont {Patrick~J}\ \bibnamefont {Coles}},\ }\bibfield  {title} {\enquote {\bibinfo {title} {Thermodynamic linear algebra},}\ }\href@noop {} {\bibfield  {journal} {\bibinfo  {journal} {arXiv preprint arXiv:2308.05660}\ } (\bibinfo {year} {2023})}\BibitemShut {NoStop}%
\bibitem [{\citenamefont {Duffield}\ \emph {et~al.}(2023)\citenamefont {Duffield}, \citenamefont {Aifer}, \citenamefont {Crooks}, \citenamefont {Ahle},\ and\ \citenamefont {Coles}}]{duffield2023thermodynamic}%
  \BibitemOpen
  \bibfield  {author} {\bibinfo {author} {\bibfnamefont {Samuel}\ \bibnamefont {Duffield}}, \bibinfo {author} {\bibfnamefont {Maxwell}\ \bibnamefont {Aifer}}, \bibinfo {author} {\bibfnamefont {Gavin}\ \bibnamefont {Crooks}}, \bibinfo {author} {\bibfnamefont {Thomas}\ \bibnamefont {Ahle}}, \ and\ \bibinfo {author} {\bibfnamefont {Patrick~J}\ \bibnamefont {Coles}},\ }\bibfield  {title} {\enquote {\bibinfo {title} {Thermodynamic matrix exponentials and thermodynamic parallelism},}\ }\href@noop {} {\bibfield  {journal} {\bibinfo  {journal} {arXiv preprint arXiv:2311.12759}\ } (\bibinfo {year} {2023})}\BibitemShut {NoStop}%
\bibitem [{\citenamefont {Hylton}(2020)}]{hylton2020thermodynamic}%
  \BibitemOpen
  \bibfield  {author} {\bibinfo {author} {\bibfnamefont {Todd}\ \bibnamefont {Hylton}},\ }\bibfield  {title} {\enquote {\bibinfo {title} {Thermodynamic neural network},}\ }\href {\doibase 10.3390/e22030256} {\bibfield  {journal} {\bibinfo  {journal} {Entropy}\ }\textbf {\bibinfo {volume} {22}},\ \bibinfo {pages} {256} (\bibinfo {year} {2020})}\BibitemShut {NoStop}%
\bibitem [{\citenamefont {Ganesh}(2017)}]{ganesh2017thermodynamic}%
  \BibitemOpen
  \bibfield  {author} {\bibinfo {author} {\bibfnamefont {Natesh}\ \bibnamefont {Ganesh}},\ }\bibfield  {title} {\enquote {\bibinfo {title} {A thermodynamic treatment of intelligent systems},}\ }in\ \href {\doibase 10.1109/ICRC.2017.8123676} {\emph {\bibinfo {booktitle} {2017 IEEE International Conference on Rebooting Computing (ICRC)}}}\ (\bibinfo {year} {2017})\ pp.\ \bibinfo {pages} {1--4}\BibitemShut {NoStop}%
\bibitem [{\citenamefont {Lipka-Bartosik}\ \emph {et~al.}(2023)\citenamefont {Lipka-Bartosik}, \citenamefont {Perarnau-Llobet},\ and\ \citenamefont {Brunner}}]{lipka2023thermodynamic}%
  \BibitemOpen
  \bibfield  {author} {\bibinfo {author} {\bibfnamefont {Patryk}\ \bibnamefont {Lipka-Bartosik}}, \bibinfo {author} {\bibfnamefont {Mart{\'\i}}\ \bibnamefont {Perarnau-Llobet}}, \ and\ \bibinfo {author} {\bibfnamefont {Nicolas}\ \bibnamefont {Brunner}},\ }\bibfield  {title} {\enquote {\bibinfo {title} {Thermodynamic computing via autonomous quantum thermal machines},}\ }\href@noop {} {\bibfield  {journal} {\bibinfo  {journal} {arXiv preprint arXiv:2308.15905}\ } (\bibinfo {year} {2023})}\BibitemShut {NoStop}%
\bibitem [{\citenamefont {Camsari}\ \emph {et~al.}(2019)\citenamefont {Camsari}, \citenamefont {Sutton},\ and\ \citenamefont {Datta}}]{Camsari_2019}%
  \BibitemOpen
  \bibfield  {author} {\bibinfo {author} {\bibfnamefont {Kerem~Y.}\ \bibnamefont {Camsari}}, \bibinfo {author} {\bibfnamefont {Brian~M.}\ \bibnamefont {Sutton}}, \ and\ \bibinfo {author} {\bibfnamefont {Supriyo}\ \bibnamefont {Datta}},\ }\bibfield  {title} {\enquote {\bibinfo {title} {p-bits for probabilistic spin logic},}\ }\href {\doibase 10.1063/1.5055860} {\bibfield  {journal} {\bibinfo  {journal} {Appl. Phys. Rev.}\ }\textbf {\bibinfo {volume} {6}},\ \bibinfo {pages} {011305} (\bibinfo {year} {2019})}\BibitemShut {NoStop}%
\bibitem [{\citenamefont {Chowdhury}\ \emph {et~al.}(2023)\citenamefont {Chowdhury}, \citenamefont {Grimaldi}, \citenamefont {Aadit}, \citenamefont {Niazi}, \citenamefont {Mohseni}, \citenamefont {Kanai}, \citenamefont {Ohno}, \citenamefont {Fukami}, \citenamefont {Theogarajan}, \citenamefont {Finocchio} \emph {et~al.}}]{chowdhury2023full}%
  \BibitemOpen
  \bibfield  {author} {\bibinfo {author} {\bibfnamefont {Shuvro}\ \bibnamefont {Chowdhury}}, \bibinfo {author} {\bibfnamefont {Andrea}\ \bibnamefont {Grimaldi}}, \bibinfo {author} {\bibfnamefont {Navid~Anjum}\ \bibnamefont {Aadit}}, \bibinfo {author} {\bibfnamefont {Shaila}\ \bibnamefont {Niazi}}, \bibinfo {author} {\bibfnamefont {Masoud}\ \bibnamefont {Mohseni}}, \bibinfo {author} {\bibfnamefont {Shun}\ \bibnamefont {Kanai}}, \bibinfo {author} {\bibfnamefont {Hideo}\ \bibnamefont {Ohno}}, \bibinfo {author} {\bibfnamefont {Shunsuke}\ \bibnamefont {Fukami}}, \bibinfo {author} {\bibfnamefont {Luke}\ \bibnamefont {Theogarajan}}, \bibinfo {author} {\bibfnamefont {Giovanni}\ \bibnamefont {Finocchio}},  \emph {et~al.},\ }\bibfield  {title} {\enquote {\bibinfo {title} {A full-stack view of probabilistic computing with p-bits: devices, architectures and algorithms},}\ }\href@noop {} {\bibfield  {journal} {\bibinfo  {journal} {IEEE Journal on Exploratory Solid-State Computational Devices and Circuits}\ } (\bibinfo
  {year} {2023})}\BibitemShut {NoStop}%
\bibitem [{\citenamefont {Misra}\ \emph {et~al.}(2023)\citenamefont {Misra}, \citenamefont {Bland}, \citenamefont {Cardwell}, \citenamefont {Incorvia}, \citenamefont {James}, \citenamefont {Kent}, \citenamefont {Schuman}, \citenamefont {Smith},\ and\ \citenamefont {Aimone}}]{misra2023probabilistic}%
  \BibitemOpen
  \bibfield  {author} {\bibinfo {author} {\bibfnamefont {Shashank}\ \bibnamefont {Misra}}, \bibinfo {author} {\bibfnamefont {Leslie~C}\ \bibnamefont {Bland}}, \bibinfo {author} {\bibfnamefont {Suma~G}\ \bibnamefont {Cardwell}}, \bibinfo {author} {\bibfnamefont {Jean Anne~C}\ \bibnamefont {Incorvia}}, \bibinfo {author} {\bibfnamefont {Conrad~D}\ \bibnamefont {James}}, \bibinfo {author} {\bibfnamefont {Andrew~D}\ \bibnamefont {Kent}}, \bibinfo {author} {\bibfnamefont {Catherine~D}\ \bibnamefont {Schuman}}, \bibinfo {author} {\bibfnamefont {J~Darby}\ \bibnamefont {Smith}}, \ and\ \bibinfo {author} {\bibfnamefont {James~B}\ \bibnamefont {Aimone}},\ }\bibfield  {title} {\enquote {\bibinfo {title} {Probabilistic neural computing with stochastic devices},}\ }\href@noop {} {\bibfield  {journal} {\bibinfo  {journal} {Advanced Materials}\ }\textbf {\bibinfo {volume} {35}},\ \bibinfo {pages} {2204569} (\bibinfo {year} {2023})}\BibitemShut {NoStop}%
\bibitem [{\citenamefont {Liu}\ \emph {et~al.}(2022)\citenamefont {Liu}, \citenamefont {Xiao}, \citenamefont {Kwon}, \citenamefont {Debusschere}, \citenamefont {Agarwal}, \citenamefont {Incorvia},\ and\ \citenamefont {Bennett}}]{liu2022bayesian}%
  \BibitemOpen
  \bibfield  {author} {\bibinfo {author} {\bibfnamefont {Samuel}\ \bibnamefont {Liu}}, \bibinfo {author} {\bibfnamefont {T~Patrick}\ \bibnamefont {Xiao}}, \bibinfo {author} {\bibfnamefont {Jaesuk}\ \bibnamefont {Kwon}}, \bibinfo {author} {\bibfnamefont {Bert~J}\ \bibnamefont {Debusschere}}, \bibinfo {author} {\bibfnamefont {Sapan}\ \bibnamefont {Agarwal}}, \bibinfo {author} {\bibfnamefont {Jean Anne~C}\ \bibnamefont {Incorvia}}, \ and\ \bibinfo {author} {\bibfnamefont {Christopher~H}\ \bibnamefont {Bennett}},\ }\bibfield  {title} {\enquote {\bibinfo {title} {Bayesian neural networks using magnetic tunnel junction-based probabilistic in-memory computing},}\ }\href@noop {} {\bibfield  {journal} {\bibinfo  {journal} {Frontiers in Nanotechnology}\ }\textbf {\bibinfo {volume} {4}},\ \bibinfo {pages} {1021943} (\bibinfo {year} {2022})}\BibitemShut {NoStop}%
\bibitem [{\citenamefont {Mansinghka}\ \emph {et~al.}(2009)\citenamefont {Mansinghka} \emph {et~al.}}]{mansinghka2009natively}%
  \BibitemOpen
  \bibfield  {author} {\bibinfo {author} {\bibfnamefont {Vikash~Kumar}\ \bibnamefont {Mansinghka}} \emph {et~al.},\ }\emph {\bibinfo {title} {Natively probabilistic computation}},\ \href@noop {} {Ph.D. thesis},\ \bibinfo  {school} {Citeseer} (\bibinfo {year} {2009})\BibitemShut {NoStop}%
\bibitem [{\citenamefont {Murphy}(2022)}]{murphy2022probabilistic}%
  \BibitemOpen
  \bibfield  {author} {\bibinfo {author} {\bibfnamefont {Kevin~P}\ \bibnamefont {Murphy}},\ }\href@noop {} {\emph {\bibinfo {title} {Probabilistic machine learning: an introduction}}}\ (\bibinfo  {publisher} {MIT press},\ \bibinfo {year} {2022})\BibitemShut {NoStop}%
\bibitem [{\citenamefont {Williams}\ and\ \citenamefont {Rasmussen}(1995)}]{williams1995gaussian}%
  \BibitemOpen
  \bibfield  {author} {\bibinfo {author} {\bibfnamefont {Christopher}\ \bibnamefont {Williams}}\ and\ \bibinfo {author} {\bibfnamefont {Carl}\ \bibnamefont {Rasmussen}},\ }\bibfield  {title} {\enquote {\bibinfo {title} {Gaussian processes for regression},}\ }\href@noop {} {\bibfield  {journal} {\bibinfo  {journal} {Advances in neural information processing systems}\ }\textbf {\bibinfo {volume} {8}} (\bibinfo {year} {1995})}\BibitemShut {NoStop}%
\bibitem [{\citenamefont {Liu}\ \emph {et~al.}(2020)\citenamefont {Liu}, \citenamefont {Lin}, \citenamefont {Padhy}, \citenamefont {Tran}, \citenamefont {Bedrax~Weiss},\ and\ \citenamefont {Lakshminarayanan}}]{liu2020simple}%
  \BibitemOpen
  \bibfield  {author} {\bibinfo {author} {\bibfnamefont {Jeremiah}\ \bibnamefont {Liu}}, \bibinfo {author} {\bibfnamefont {Zi}~\bibnamefont {Lin}}, \bibinfo {author} {\bibfnamefont {Shreyas}\ \bibnamefont {Padhy}}, \bibinfo {author} {\bibfnamefont {Dustin}\ \bibnamefont {Tran}}, \bibinfo {author} {\bibfnamefont {Tania}\ \bibnamefont {Bedrax~Weiss}}, \ and\ \bibinfo {author} {\bibfnamefont {Balaji}\ \bibnamefont {Lakshminarayanan}},\ }\bibfield  {title} {\enquote {\bibinfo {title} {Simple and principled uncertainty estimation with deterministic deep learning via distance awareness},}\ }\href@noop {} {\bibfield  {journal} {\bibinfo  {journal} {Advances in Neural Information Processing Systems}\ }\textbf {\bibinfo {volume} {33}},\ \bibinfo {pages} {7498--7512} (\bibinfo {year} {2020})}\BibitemShut {NoStop}%
\bibitem [{\citenamefont {Chen}\ \emph {et~al.}(2014)\citenamefont {Chen}, \citenamefont {Fox},\ and\ \citenamefont {Guestrin}}]{chen2014stochastic}%
  \BibitemOpen
  \bibfield  {author} {\bibinfo {author} {\bibfnamefont {Tianqi}\ \bibnamefont {Chen}}, \bibinfo {author} {\bibfnamefont {Emily}\ \bibnamefont {Fox}}, \ and\ \bibinfo {author} {\bibfnamefont {Carlos}\ \bibnamefont {Guestrin}},\ }\bibfield  {title} {\enquote {\bibinfo {title} {Stochastic gradient hamiltonian monte carlo},}\ }in\ \href@noop {} {\emph {\bibinfo {booktitle} {International conference on machine learning}}}\ (\bibinfo {organization} {PMLR},\ \bibinfo {year} {2014})\ pp.\ \bibinfo {pages} {1683--1691}\BibitemShut {NoStop}%
\bibitem [{\citenamefont {Shor}(1995)}]{shor1995scheme}%
  \BibitemOpen
  \bibfield  {author} {\bibinfo {author} {\bibfnamefont {Peter~W}\ \bibnamefont {Shor}},\ }\bibfield  {title} {\enquote {\bibinfo {title} {Scheme for reducing decoherence in quantum computer memory},}\ }\href {\doibase https://doi.org/10.1103/PhysRevA.52.R2493} {\bibfield  {journal} {\bibinfo  {journal} {Physical review A}\ }\textbf {\bibinfo {volume} {52}},\ \bibinfo {pages} {R2493} (\bibinfo {year} {1995})}\BibitemShut {NoStop}%
\bibitem [{\citenamefont {Shor}(1999)}]{shor1999polynomial}%
  \BibitemOpen
  \bibfield  {author} {\bibinfo {author} {\bibfnamefont {Peter~W}\ \bibnamefont {Shor}},\ }\bibfield  {title} {\enquote {\bibinfo {title} {Polynomial-time algorithms for prime factorization and discrete logarithms on a quantum computer},}\ }\href {\doibase 10.1137/S0036144598347011} {\bibfield  {journal} {\bibinfo  {journal} {SIAM review}\ }\textbf {\bibinfo {volume} {41}},\ \bibinfo {pages} {303--332} (\bibinfo {year} {1999})}\BibitemShut {NoStop}%
\bibitem [{\citenamefont {Chuang}\ \emph {et~al.}(1998)\citenamefont {Chuang}, \citenamefont {Gershenfeld},\ and\ \citenamefont {Kubinec}}]{chuang1998experimental}%
  \BibitemOpen
  \bibfield  {author} {\bibinfo {author} {\bibfnamefont {Isaac~L}\ \bibnamefont {Chuang}}, \bibinfo {author} {\bibfnamefont {Neil}\ \bibnamefont {Gershenfeld}}, \ and\ \bibinfo {author} {\bibfnamefont {Mark}\ \bibnamefont {Kubinec}},\ }\bibfield  {title} {\enquote {\bibinfo {title} {Experimental implementation of fast quantum searching},}\ }\href@noop {} {\bibfield  {journal} {\bibinfo  {journal} {Physical review letters}\ }\textbf {\bibinfo {volume} {80}},\ \bibinfo {pages} {3408} (\bibinfo {year} {1998})}\BibitemShut {NoStop}%
\end{thebibliography}%

\pagebreak

\begin{appendix}

\begin{center}
\Large Supplemental Information for \\``Thermodynamic Computing System for AI Applications''
\end{center}

\section{Overview}\label{secA1}

In this Supplemental Information, we will cover the following topics:
\begin{itemize}
\item Additional experimental implementations  
\item Device characterization
\item Derivation of circuit Hamiltonian
\item Details for covariance-based sampling
\end{itemize}

\section{Additional experimental implementations} 

\subsection{Additional matrix inversion data} 

\begin{figure}[h]
    \centering
    \includegraphics[width=0.47\textwidth]{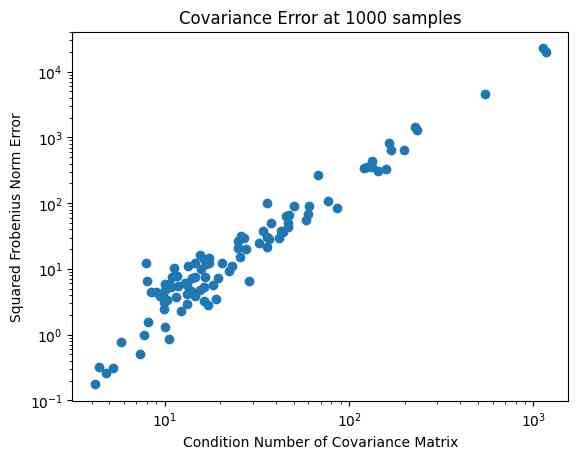}
    \includegraphics[width=0.47\textwidth]{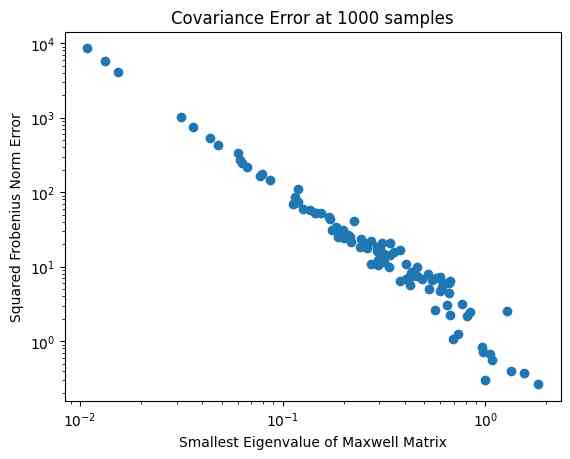}
    \caption{\justifying \textbf{Effect of condition number and smallest eigenvalue on the inversion error}. We consider matrix inversion while varying certain properties of the matrix, for symmetric positive definite matrices. The left panel shows the inversion error versus the condition number of the matrix. The right panel shows the inversion error versus the smallest eigenvalue of the matrix.}
    \label{fig:ConditionNumber}
\end{figure}

Here we provide some additional data for inverting matrices on our SPU. Figure~\ref{fig:ConditionNumber} shows the effect of certain matrix properties on the inversion error from our SPU, for the case of symmetric positive definite matrices. At the theoretical level, it is expected that increasing the condition number, $\kappa$ (the ratio of the largest to the smallest eigenvalue of the matrix), should increase the difficulty of matrix inversion. For example, for our thermodynamic matrix inversion algorithm, the runtime in the underdamped regime was shown to be $O(d^2 \kappa \epsilon^{-2})$~\cite{aifer2023thermodynamic}. The left panel of Fig.~\ref{fig:ConditionNumber} indeed shows that error rises with condition number, as expected. Similarly, we see in the right panel of Fig.~\ref{fig:ConditionNumber} that the error goes down as the smallest eigenvalue increases in magnitude.

\subsection{Least-squares regression} 

Here we consider an additional application for our thermodynamic hardware: least-squares regression. Specifically, linear least-squares involves matrix inversion as a subroutine. Hence, we can run this subroutine on our hardware, and when appropriately scaled up, our hardware could accelerate this subroutine of the least-squares regression algorithm. 

In more detail, consider a dataset where $y$ is a vector whose $i$th element is the $i$th observation of the dependent variable, and where $X$ is a matrix whose $(i,j)$th element is the $i$th observation of the $j$th dependent variable. Then the solution of the linear least-squares regression problem is given by the following equation: 
\begin{equation}\label{eqn_LinearRegression}
    \beta = (X^{T}X)^{-1} y
\end{equation}
Here, $\beta$ is the vector of unknown parameters, and the above equation gives the optimal set of parameters. Here there are two parameters, the intercept $\beta_0$ and the slope $\beta_1$.

\begin{figure}
    \centering
    \includegraphics[width=0.48\textwidth]{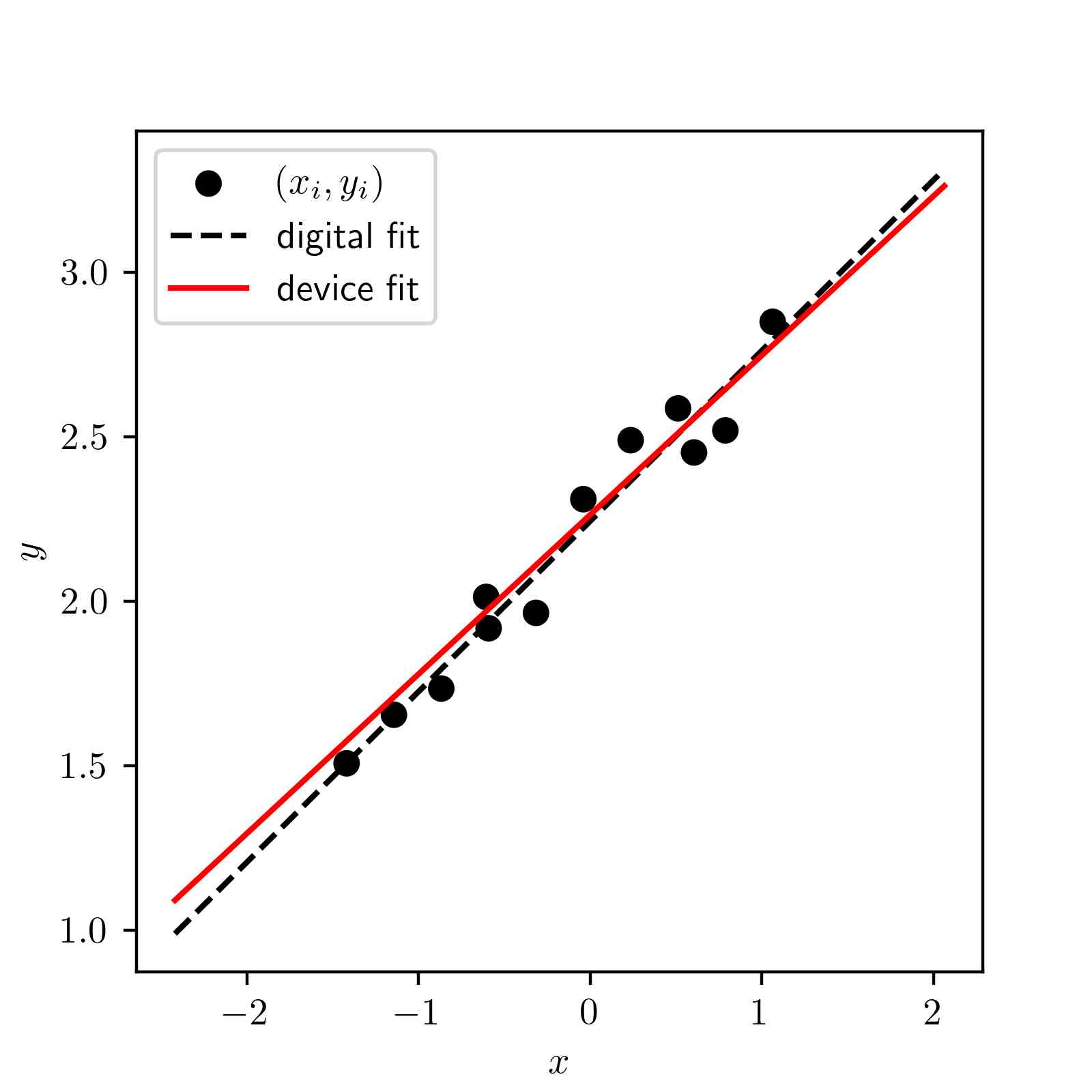}
    \caption{\justifying \textbf{Running the linear least-squares regression problem on our SPU}. The training data are shown as blue dots. The results from the SPU are shown in the solid red curve, while the results from a digital device are shown as the black dotted line. For the SPU case, the matrix inversion portion of Eq.~\eqref{eqn_LinearRegression} are run on the SPU. We find the $\beta^{\mathrm{SPU}}_0, \beta^{\mathrm{SPU}}_1 = (2.26, 0.49)$ and $\beta^{\mathrm{digital}}_0, \beta^{\mathrm{digital}}_1 = (2.24, 0.52)$.}
\label{fig:LinearRegression}
\end{figure}

Figure~\ref{fig:LinearRegression} shows the results of applying our SPU to a linear least-squares regression problem for a particular dataset. Namely, we use our SPU to compute the matrix inverse that appears in Eq.~\eqref{eqn_LinearRegression}, and then the matrix-vector multiplication in this equation is performed digitally. One can see that a reasonable fit to the data is achieved with our SPU, fairly similar to the result obtained from performing the whole regression algorithm digitally.

\section{Device Characterization}

\subsection{Spectroscopy with Noise Driving}

The characterization of the device means to identify the values of the electrical components as they are in the final product (how much they differ from the nominal design values). One method that enables us to do this without modifying the PCB or requiring outside equipment is noise-driven spectroscopy. 

This technique involves driving our system with FPGA white noise and monitoring the response of the system by measuring the voltage at the output. If we assume that the driving noise is Gaussian white noise, the power spectrum of the noise is flat and has infinite bandwidth. This means that the system is effectively being driven at every frequency, and as such the output will contain the response of the system to all frequencies and should contain the resonance frequency peak of the RLC unit cell. The noise used in the device is pseudo-Gaussian white noise that has a relatively flat power spectrum up to the sampling frequency of the device of 12 MHz. This is not a limitation of this method because we use a similar digital pseudo-random noise source for our model.

\begin{figure}
    \centering
    \includegraphics[width=0.7\textwidth]{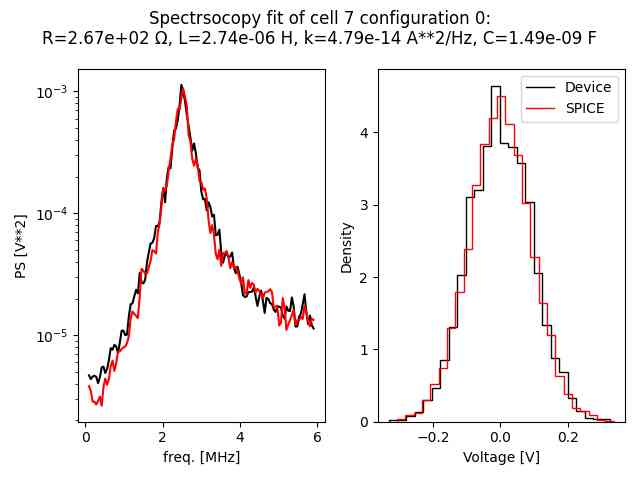}
    \caption{\justifying Example fit of the power spectrum (Left) and histogram (Right) from cell 7 of the device.}
    \label{fig:spectroscopy_fit}
\end{figure}

By running the same protocol on a simple SPICE model of the device, we are able to fit and determine the values of the circuit parameters. The model used is a the one from fig.~\ref{fig:PCB_AllAll}, with fixed components and a Gaussian current source. We run a SPICE simulation to gather samples in the same conditions of the device, then we determine the power spectrum and fit to the device. The cost function of the fit includes the spectrum error and the variance error (the difference between the device sample variance and the SPICE simulation sample variance). Figure~\ref{fig:spectroscopy_fit} shows a visualization of the fit results, we see great agreement between the SPICE model and the device. The best fit values were relatively close to the nominal values except for non-uniformity due to the specific implementation of the transformer-based coupling used (see Supplemental Information).

A similar approach can be used to determine the effective capacitive coupling between the cells, with an added penalty term in the cost function related to the covariance between the cells. The best fit values for the capacitive coupling can be up to a factor of two smaller than the nominal values. This can be attributed to the resistive damping within the coupling circuitry that was not included in this simplified model. This effective lowering of the coupling can be accounted for by using the best fit values in the compilation or embedding of the desired covariance matrix. 

Table~\ref{tab:fit_params_config3} shows the fit parameters for each cell at maximum capacitance (3 in the FPGA configuration matrix).

\begin{table}
\centering
\caption{Table of best fit parameters from all 8 cells in configuration 3 of the unit cell capacitor banks.}
\begin{tabular}{c|c|c|c|c}
    \hline
    Cell ID &  $L (\mu \mathrm{H})$ & $R (\Omega)$ & $\kappa_0 (\mu \mathrm{A^2 Hz^{-1}})$ & $C (\mathrm{nF})$ \\ [0.5ex] 
    \hline\hline
     7 & 1.77 & 76.7 & 0.206 & 6.77 \\
     6 & 2.22 & 94.6 & 0.0632 & 6.14 \\
     5 & 1.48 & 63.3 & 0.0738 & 6.22 \\
     4 & 1.32 & 30.5 & 0.114 & 6.44 \\
     3 & 1.29 & 43.7 & 0.0631 & 6.21 \\
     2 & 1.19 & 34.8 & 0.0448 & 6.71 \\
     1 & 0.989 & 30.7 & 0.0660 & 6.44 \\
     0 & 0.918 & 26.9 & 0.0697 & 6.40 \\ [1ex] 
    \hline
\end{tabular}
\label{tab:fit_params_config3}
\end{table}

\subsubsection{Spectroscopy-based method for finding hardware issues}

When assembling a PCB, there is always a possibility, however small, of hardware issues. In order to quickly characterize the device for any faults, a two-cell spectroscopy experiment was used to determine the qualitative function. In this experiment, one cell is used as the drive cell while the second cell is used as a probe cell. The drive cell is the only cell with any noise driving (usually a noise level of 100\%), while the probe cell is where we take voltage measurements (samples). When the drive cell and the probe cell are the same, this means we are looking at the self-driving of that cell. This is then repeated for all combinations of cells and for the cases of zero coupling, positive coupling, and negative coupling. The samples from the probe cell are then used to calculate the power spectrum. The power spectrum of each cell and each coupling should have a peak at the designed frequency, if this peak is absent, below a certain threshold or shifted considerably in frequency, then we can flag this for further investigation. Running this test on the device takes on the order of 5 minutes and can pin-point problems that would have taken hours by hand.

\subsection{Post-processing to correct non-uniformity}

The current design of the PCB has an unfortunate property that makes the unit cells have different properties when they should be identical. For example, when setting up the device to have the identity covariance matrix, the covariance matrix of the samples does not have a uniform diagonal as would be expected. Specifically, the variances are monotonically increasing with the cell number; cell 0 has a smaller variance than cell 7 even when both cells have the same capacitance.

\begin{figure}
    \centering
    \includegraphics[width=0.6\textwidth]{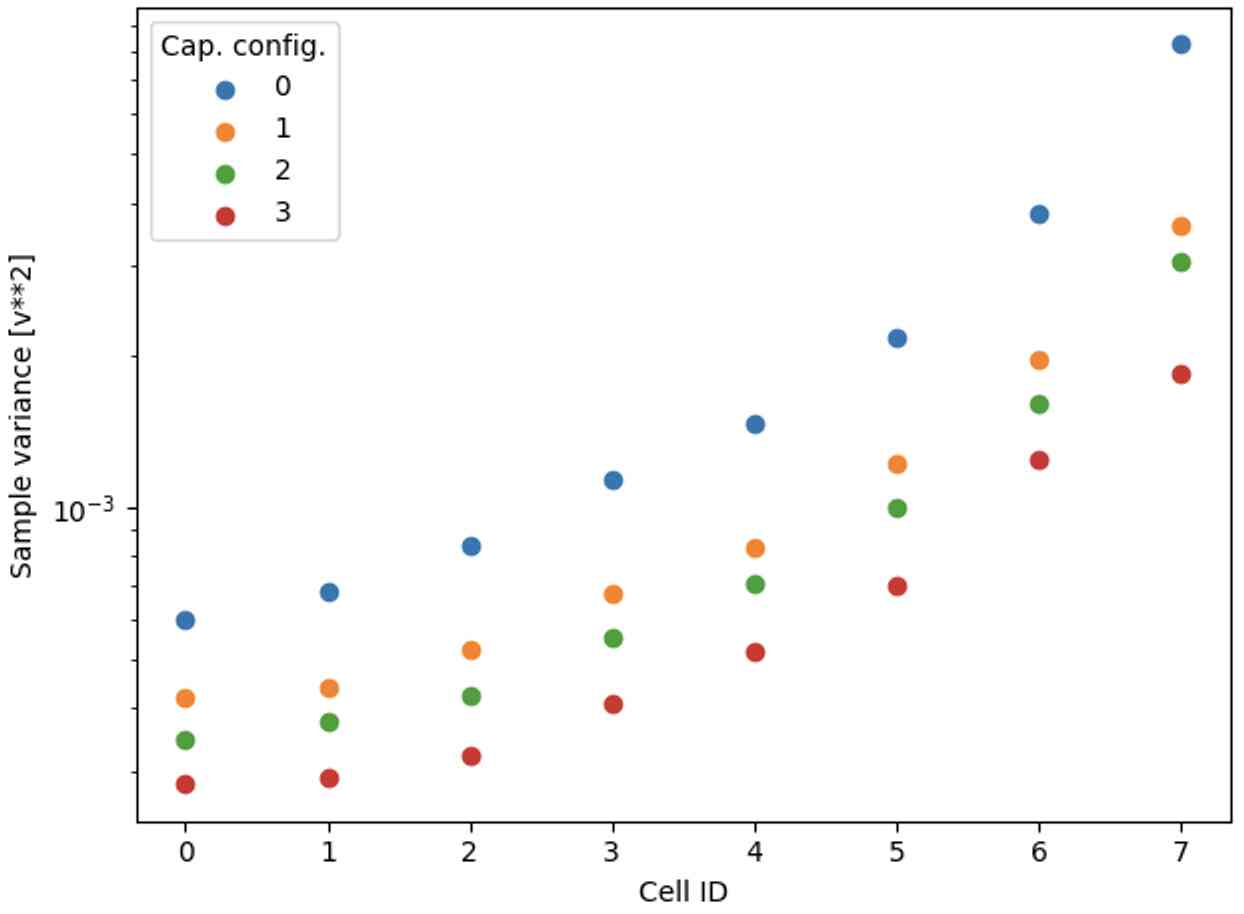}
    \caption{\justifying Sample variance of each cell in different capacitor bank configurations.}
    \label{fig:SampleVarVsCell1}
\end{figure}

Figure~\ref{fig:SampleVarVsCell1} shows a plot of the sample variance for each cell for each capacitor configuration (the configuration of the switching capacitor bank). In these plotted examples, the capacitance matrix is diagonal (all coupling turned off) and uniform for all cells.

As discussed in the main text, the form of the sample covariance matrix, assuming that both the resistance matrix and the noise driving are proportional to the identity, should be given by
\begin{equation}\label{eq:sample_cov_simple}
\Sigma_\mathrm{V} = \kappa_0 R C^{-1}
\end{equation}

This theory predicts that the variance of the samples from the device with an identity capacitance matrix should be identity. This is not what we see in practice, as shown in Figure~\ref{fig:SampleVarVsCell1}.

\subsubsection{Calibration method}

In an attempt to overcome this issue, we use the insight from the analytical treatment to correct for this through post-processing. 
\begin{enumerate}
    \item The first step is to take samples from the identity capacitance matrix as a baseline (usually configuration 0 of the capacitor bank in the unit cells, with all couplings turned off).
    \item The second step is to calculate a scaling vector based on the identity samples. This scaling vector is calculated from the diagonal values of the sample covariance matrix from step 1. The scaling vector is given by the inverse of the square-root of the diagonal.
    \item The third step is to take samples from the desired experiment with arbitrary capacitance matrix.
    \item The fourth step is to re-scale the samples from the desired experiment with the scaling vector found from step 2. This is done by multiplying the samples from each cell by the corresponding value in the scaling vector.
\end{enumerate}

\begin{figure}
    \centering
    \includegraphics[width=0.7\textwidth]{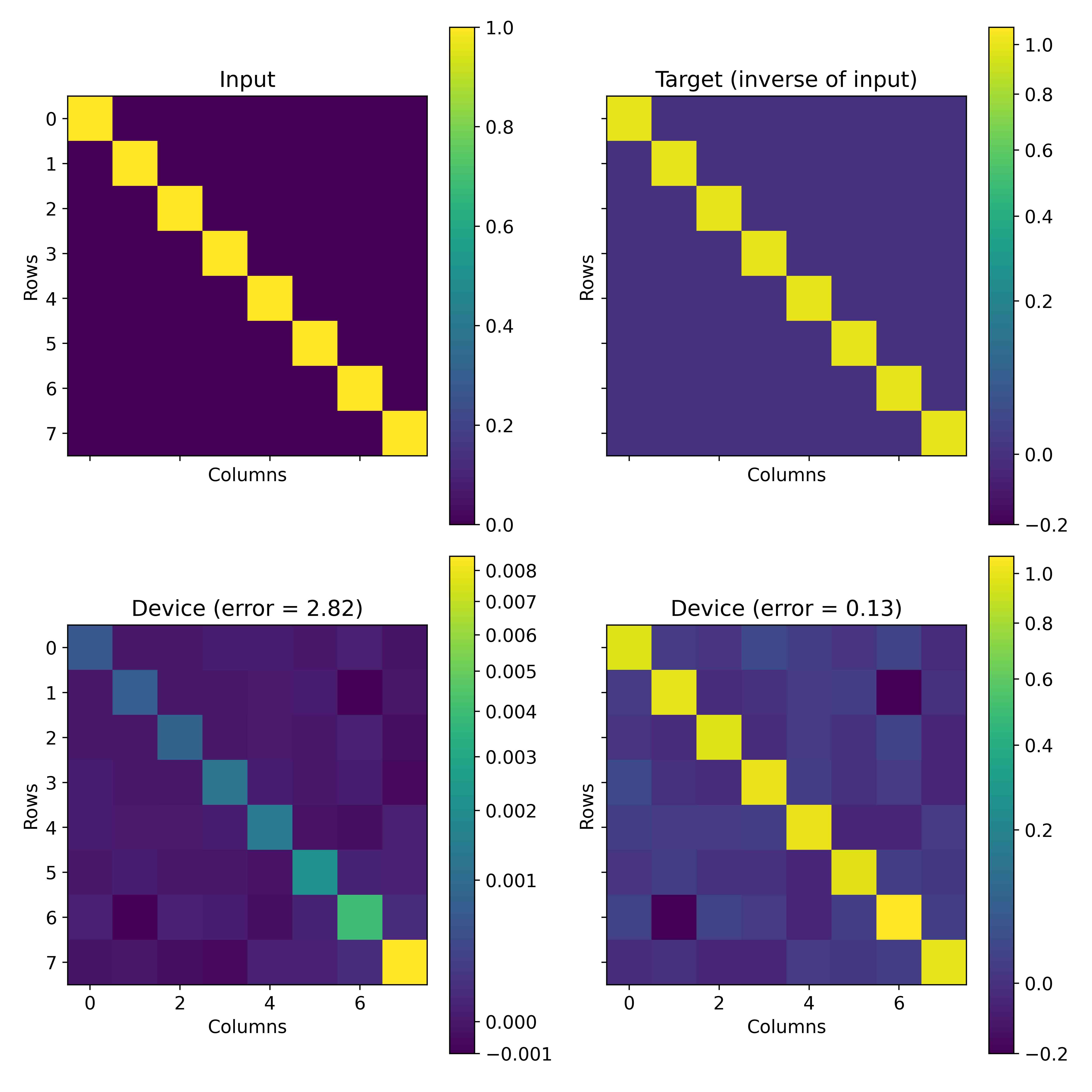}
    \caption{\justifying Heat maps of the input capacitance matrix (Top Left), the target covariance matrix, which corresponds to the inverse of the input (Top Right), the raw uncorrected sample covariance matrix (Bottom Left), and the corrected sample covariance matrix (Bottom Left).}
    \label{fig:IdentityMatrix_Processed}
\end{figure}

Doing this on the identity matrix itself (configuration 0 of the capacitor banks with all couplings turned off) results in the plots shown in Figure~\ref{fig:IdentityMatrix_Processed}, where the non-uniformity of the diagonal is corrected.

\subsubsection{Model based on parallel loading}

\begin{figure}
    \centering
    \includegraphics[width=0.5\textwidth]{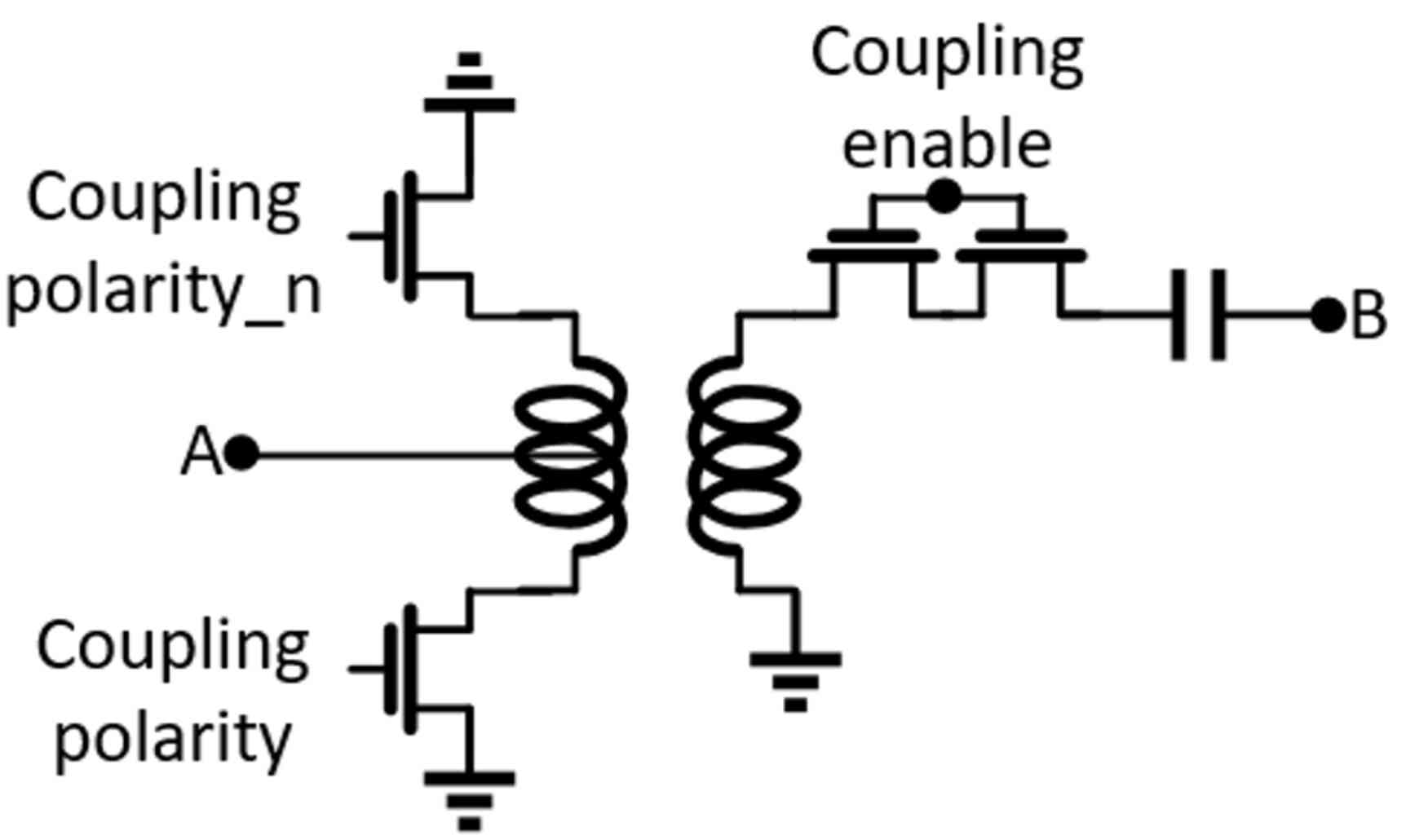}
    \caption{\justifying Circuit diagram of the coupling circuit responsible for connecting all the cells.}
    \label{fig:BipolarCoupling}
\end{figure}

The explanation behind this non-uniform variance is the non-symmetric implementation of the coupling unit on the PCB. The coupling unit circuit diagram is shown in Figure~\ref{fig:BipolarCoupling}.

This circuit is inserted between every pair of cells on the PCB to act as bipolar capacitive coupling. In Fig.~\ref{fig:BipolarCoupling}, one cell is connected to node ‘A’ and a second cell is connected to node ‘B’. In this configuration, cell ‘A’ is directly connected to the transformer while cell ‘B’ is indirectly connected to it through a capacitor. This has the effect of asymmetrically loading the cells.

This coupling circuit is identical for all couplings on the PCB and is always in the same orientation. This means that the cells on the left half of the PCB are connected  predominantly to node ‘A’, while cells on the right half of the PCB are predominantly connected to node ‘B’. This means that cells with a higher proportion of their connections being on node 'B' will have less parallel damping from the transformers than cells predominantly connected to node 'A'. When numbering the cells from left to right, we get that the lower numbered cells have more effective loading than the higher numbered ones.

To test this theory, we devise a simple model based on parallel loading of the added resistance by the coupling circuit. The formula for the effective resistance from two parallel resistors is given by

\begin{equation}
    \frac{1}{R_{eff}} = \frac{1}{R_1} + \frac{1}{R_2}
\end{equation}

This can be extended to our case by introducing the cell resistance $R$, the coupling circuit resistance $R_\mathrm{c}$, the total number of cells in the circuit $N_\mathrm{tot}$, and the individual cell ID $i$. The equation for the effective resistance seen by cell $i $ in the PCB becomes

\begin{equation}
    R_{i}' = \frac{1}{\frac{1}{R} + \frac{N_\mathrm{tot} - 1-i}{R_\mathrm{c}}}.
\end{equation}

On first approximation from Eq.~\ref{eq:sample_cov_simple}, the covariance matrix should be directly proportional to the effective resistance in the cells (for a constant capacitance value). We can thus suggest the following model for the variances and fit the data (where $N_\mathrm{tot} = 8$)

\begin{equation}
    \Sigma_{ii} = \frac{a b}{b + (7-i)a}.
\end{equation}

\begin{figure}
    \centering
    \includegraphics[width=0.6\textwidth]{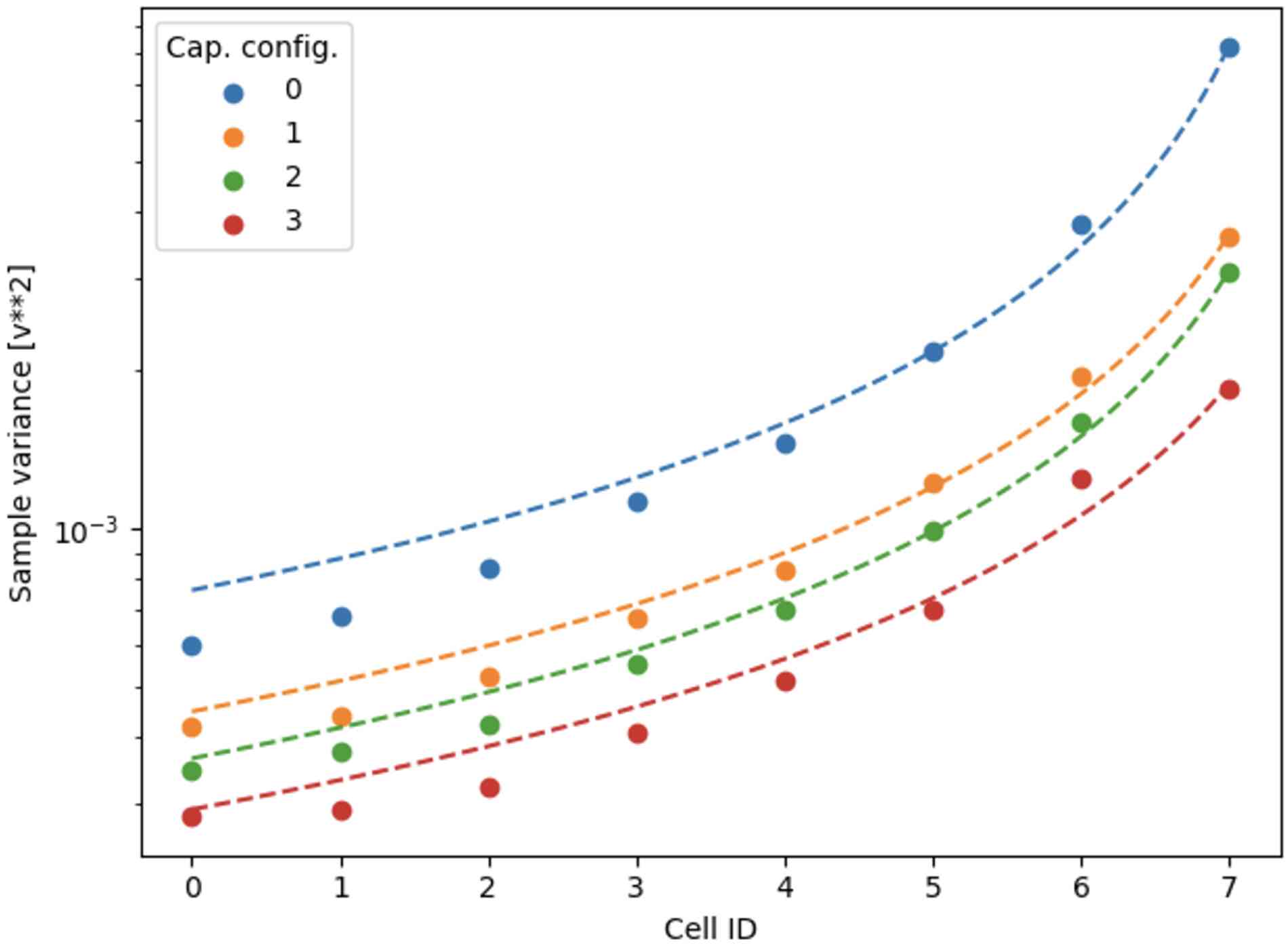}
    \caption{\justifying Fit of the parallel resistive loading model (dashed lines) to the sample variance data (solid points).}
    \label{fig:SampleVar_Fit}
\end{figure}

Figure~\ref{fig:SampleVar_Fit} shows this model fit to the device sample variances, which appears to explain the main trend. In future versions of the PCB, this feature could be corrected by including a capacitor on both sides of the transformer for example.

\section{Derivation of capacitively coupled oscillators Hamiltonian}

The stationary distribution of a system undergoing Langevin dynamics will be closely related to the noiseless and dissipation-free Hamiltonian of the same system. In this section we derive a simplified Hamiltonian that describes well our device in the limit of zero dissipation and zero noise.

\subsection{Circuit Lagrangian}

\begin{figure}[h]
    \centering
    \includegraphics[width=0.7\columnwidth]{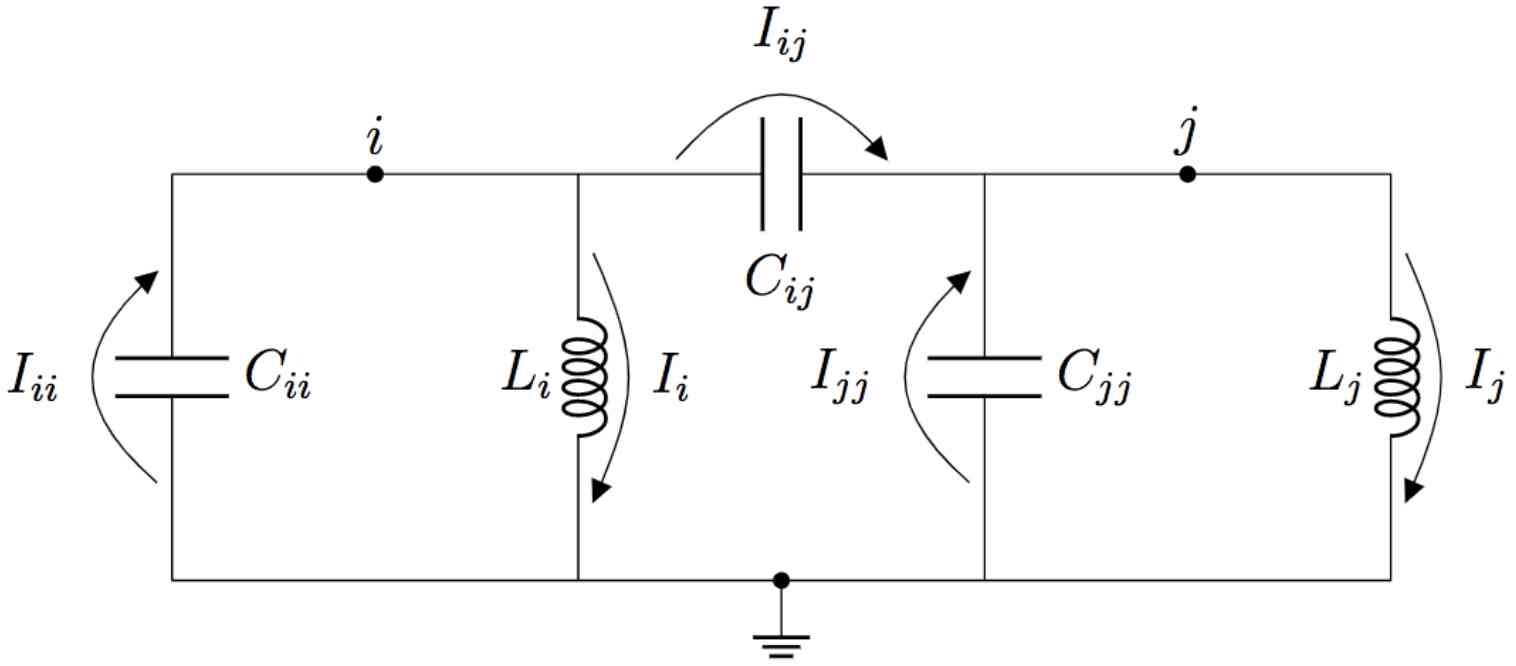}
    \caption{\justifying Circuit diagram for a pair of capacitively coupled oscillators in a network of capacitively coupled harmonic oscillators that maps to a multivariate normal probability distribution.}
    \label{fig:TwoCapCoupledCells}
\end{figure}

Consider a set of $N$ harmonic oscillators, each capacitively coupled to all the others.  Any pair $i,j$ of oscillators can be represented by the subcircuit shown in Figure~\ref{fig:TwoCapCoupledCells}.

The total inductive energy of the circuit is
\[
E_L = \frac{1}{2}\sum_{i=1}^N L_i I_i^2
\]
If we define the inductive matrix as
\[
\mathbf{L}_{kl} = \begin{cases}
L_{k} & \text{if } k=l \\
0 &  \text{if } k\neq l \\
\end{cases},
\]
then we can write the total inductive energy as
\[
E_L = \frac{1}{2}\vec{I}^T \mathbf{L} \vec{I}.
\]
The total capacitive energy is
\[
E_C = \frac{1}{2}\sum_{i=1}^N \left[C_{ii}V_i^2 + \sum_{j=i+1}^N C_{ij}\left(V_i - V_j\right)^2 \right].
\]
To simplify the subsequent calculus, define the Maxwell capacitance matrix $\mathbf{C}$ as
\[
\mathbf{C}_{kl} = \begin{cases}
\sum_j C_{kj} & \text{if } k=l \\
- C_{kl} &  \text{if } k\neq l \\
\end{cases}.
\]
Then we can rewrite the capacitive energy as
\[
E_C = \frac{1}{2} \sum_{ij} V_i \mathbf{C}_{ij} V_j.
\]
Suppose we want to treat the currents through the inductors as our position coordinates.  To write down a Lagrangian, we rewrite the voltages in terms of the time derivatives of these position coordinates.  For each voltage, we have the relation $V_i = L_i \dot{I}_i$.  Thus we can write the capacitive energy as
\[
E_C = \frac{1}{2} \sum_{ij} \dot{I}_i L_i \mathbf{C}_{ij} L_j \dot{I}_j.
\]
Now the Lagrangian for the system can be written as
\[
\mathcal{L}(\{I_i, \dot{I}_i\}_1^N) = \frac{1}{2} \sum_{ij} \dot{I}_i L_i \mathbf{C}_{ij} L_j \dot{I}_j - \frac{1}{2}\sum_{i=1}^N L_i I_i^2.
\]

\subsection{Circuit Hamiltonian}
Hamiltonians are defined in terms of $N$ positions $q_i$, $N$ momenta $p_i$, and Lagrangian $\mathcal{L}$ as
\[
\mathcal{H} = \sum_i p_i \dot{q}_i - \mathcal{L}
\]
where the momenta are defined as
\[
p_i = \frac{\partial \mathcal{L}(\{q_j, \dot{q}_j\}_1^N)}{\partial \dot{q}_i}
\]
In our case the momentum is
\begin{align*}
p_k &= \frac{\partial \mathcal{L}(\{I_j, \dot{I}_j\}_1^N)}{\partial \dot{I}_k}
\\&= \frac{1}{2}\sum_{ij} \left(\delta_{ik}L_k \mathbf{C}_{kj} L_j \dot{I}_j + \delta_{jk}\dot{I}_iL_i \mathbf{C}_{ik}L_k\right)
\\&= \frac{1}{2}\sum_{j} L_k \mathbf{C}_{kj} L_j \dot{I}_j + \sum_{j}\dot{I}_jL_j \mathbf{C}_{jk}L_k
\\&= \sum_{j} L_k \mathbf{C}_{kj} L_j \dot{I}_j
\end{align*}
where, going line by line, we used: definition of conjugate momentum; product rule for derivatives; contract kronecker delta and relabel indices; and $\mathbf{C}$ is a symmetric matrix.  We can summarize this in a matrix equation:
\[
\vec{p} = \mathbf{L}\mathbf{C}\mathbf{L}\vec{\dot{I}}
\]
implying
\[
\vec{\dot{I}} = \mathbf{L}^{-1}\mathbf{C}^{-1}\mathbf{L}^{-1}\vec{p}.
\]

Then the Lagrangian can be rewritten as
\begin{align*}
\mathcal{L}(\{I_i, p_i\}_1^N) &= \frac{1}{2} \vec{\dot{I}}^T \mathbf{L} \mathbf{C} \mathbf{L} \vec{\dot{I}} - \frac{1}{2}\vec{I}^T \mathbf{L} \vec{I}
\\&= \frac{1}{2} \vec{p}^T \mathbf{L}^{-1}\mathbf{C}^{-1}\mathbf{L}^{-1}\mathbf{L} \mathbf{C} \mathbf{L} \mathbf{L}^{-1}\mathbf{C}^{-1}\mathbf{L}^{-1}\vec{p} - \frac{1}{2}\vec{I}^T \mathbf{L} \vec{I}
\\&= \frac{1}{2} \vec{p}^T\mathbf{L}^{-1}\mathbf{C}^{-1}\mathbf{L}^{-1}\vec{p} - \frac{1}{2}\vec{I}^T \mathbf{L} \vec{I}.
\end{align*}
This implies the Hamiltonian of the system is
\[
\mathcal{H} = \frac{1}{2}\vec{p}^T\mathbf{L}^{-1}\mathbf{C}^{-1}\mathbf{L}^{-1}\vec{p} + \frac{1}{2}\vec{I}^T \mathbf{L} \vec{I}.
\]

In terms of circuit parameters, we have the voltage across the inductor as
\[
\vec{V}_{\text{ind}} = \mathbf{L}\vec{\dot{I}},
\]
so the momentum is
\[
\vec{p} = \mathbf{L}\mathbf{C}\vec{V}_{\text{ind}}
\]
and so the voltage in terms of the momentum is
\[
\vec{V}_{\text{ind}} = \mathbf{C}^{-1}\mathbf{L}^{-1}\vec{p}
\]
Then, the Hamiltonian can also be calculated as
\[
\mathcal{H} = \frac{1}{2}\vec{V}_{\text{ind}}^{T}\mathbf{C}\vec{V}_{\text{ind}} + \frac{1}{2}\vec{I}^T \mathbf{L} \vec{I}.
\]

\subsection{Change of coordinates}
The correspondence to Langevin dynamics is simplified by introducing the following change of variables. First define the ``Maxwell charge vector" as 
\begin{equation}
\label{maxwell-charge-def}
    \mathcal{Q} = \bold{C} V,
\end{equation}
and the ``flux vector" as
\begin{equation}
\label{flux-def}
    \Phi = \bold{L} I.
\end{equation}
Physically the components of the Maxwell charge vector do not correspond to the charges on individual capacitors, but satisfy the relation $\dot{\mathcal{Q}}=-I$, where the negative sign is due to the convention we have chosen for current. In terms of these variables, the Hamiltonian is
\begin{align}\label{eq:hamiltonian-Q-phi}
    \mathcal{H} \left(\Phi, \mathcal{Q}\right) &=   \frac{1}{2}\Phi^T \mathbf{L}^{-1} \Phi + \frac{1}{2}\mathcal{Q}^T\mathbf{C}^{-1}\mathcal{Q}.
\end{align}
For a noiseless system with this Hamiltonian, the dynamics would be given by Hamilton's equations
\begin{align}
    \dot{\Phi} &= \nabla_\mathcal{Q} \mathcal{H} = \bold{C}^{-1}\mathcal{Q}\\
    \dot{\mathcal{Q}} &= -\nabla_\Phi \mathcal{H} = -L^{-1}\Phi,
\end{align}
which is equivalent to the dynamical equations $\dot{\mathcal{Q}} = -I$ and $\dot{\Phi} = V$, confirming that $\Phi$ and $\mathcal{Q}$ are canonically conjugate variables. It is clear from our statement of Hamilton's equations that $\Phi$ is playing the role of position here, and $\mathcal{Q}$ is playing the role of momentum.

\section{Covariance-based sampling}

As explained in the main text, a coordinate transformation is made to the canonically conjugate vectors $\Phi$ and $\mathcal{Q}$, 
\begin{equation}
    \Phi = L I,  \: \: \: \: \: \: \: \mathcal{Q} = \bold{C} V.
\end{equation}
The voltage vector $V$ and current vector $I$ evolve under the following equations, which were derived earlier

\begin{align}\label{eq:I-langevin-app}
    \dif I &= \mathbf{L}^{-1} V \dif t \\
    \label{eq:V-langevin-app}
    \dif V &= -\mathbf{C}^{-1} \mathbf{R}^{-1} V \dif t - \mathbf{C}^{-1} I \dif t + \sqrt{2\kappa_0}\mathbf{C}^{-1} \mathcal{N}[0,\mathbb{I}\,\dif t].
\end{align}
In terms of $\Phi$ and $\mathcal{Q}$, Eqs. \eqref{eq:I-langevin-app} and \eqref{eq:V-langevin-app} can be written as
\begin{align}
\label{eq:langevin-phi-app}
    \dif \Phi &= \bold{C}^{-1}\mathcal{Q} \, \dif t \\
    \label{eq:langevin-Q-app}
    \dif \mathcal{Q} &= -  L^{-1}\Phi\, \dif t - R^{-1} \bold{C}^{-1}\mathcal{Q}\, \dif t + \mathcal{N}[0,2R^{-1}\beta^{-1}\mathbb{I}\,\dif t].
\end{align}
Note that in the absence of noise and damping, the equations of motion would be
\begin{align}
\label{eq:noiseless-phi-app}
    \dif \Phi &= \bold{C}^{-1}\mathcal{Q} \, \dif t \\
    \label{eq:noiseless-Q-app}
    \dif \mathcal{Q} &= -  L^{-1}\Phi\, \dif t.
\end{align}
These are identical to Hamilton's equations when the Hamiltonian function is given by
\begin{equation}
    \mathcal{H}(\Phi, \mathcal{Q}) = \frac{1}{2} \mathcal{Q}^\intercal \bold{C}^{-1}  \mathcal{Q}  + \frac{1}{2} \Phi^\intercal L^{-1} \Phi.
\end{equation}
If we identify $\Phi$ as the position coordinate and $\mathcal{Q}$ as the momentum coordinate. That is,
\begin{equation}
    \dot{\Phi} = \partial_{\mathcal{Q}}\mathcal{H},
\end{equation}
and
\begin{equation}
    \dot{\mathcal{Q}} = -\partial_{\Phi}\mathcal{H},
\end{equation}
meaning that $\Phi$ and $\mathcal{Q}$ are canonically conjugate variables.

It is clear that Eqs.~\eqref{eq:langevin-phi-app} and \eqref{eq:langevin-Q-app} are equivalent to \eqref{eq:UDL-x} and \eqref{eq:UDL-p} when we make the identifications $x = \Phi$, $p = \mathcal{Q}$, $M = \bold{C}$, $\gamma = R^{-1}$, and $U(x) = U\left(\Phi\right) = \frac{1}{2}\Phi^T \bold{L}^{-1} \Phi$. In these coordinates the Hamiltonian, without noise or dissipation, of the system is expressed as
\begin{align}\label{eq:hamiltonian}
    \mathcal{H} \left(\Phi, \mathcal{Q}\right) &=   \frac{1}{2}\Phi^T \mathbf{L}^{-1} \Phi + \frac{1}{2}\mathcal{Q}^T\mathbf{C}^{-1}\mathcal{Q},
\end{align}
Because the Hamiltonian takes this form, it follows from the results of \cite{chen2014stochastic} that the Langevin equation has the stationary distribution quoted in the main text
\begin{equation}
    \label{eq:gibbs-Q-Phi-app}
    \Phi \sim \mathcal{N}[0,\beta^{-1}\bold{L}], \: \: \: \: \: \: \mathcal{Q} \sim \mathcal{N}[0,\beta^{-1}\bold{C}].
\end{equation}
Note that the charge vector $\mathcal{Q}$ can be obtained by integrating the current that flows through each cell, that is
\begin{equation}
\mathcal{Q}_j(t) = \int_0^t dt' I_j(t'),
\end{equation}
where $I_j$ is the current flowing through the resistor in the $i$th cell. These integrals can be carried out by sending the voltage across each resistor into an analog integrator circuit, and measuring the integrated charge to obtain $\mathcal{Q}$.

\end{appendix}

\end{document}